\documentclass[12pt]{article}

\usepackage{array} 
\usepackage{amssymb}
\usepackage{graphics,graphpap}
\usepackage{graphicx}
\usepackage{color}
\usepackage{graphicx}
\usepackage{dcolumn}
\usepackage{epsfig}
\usepackage{epstopdf}
\usepackage{amsmath}
\usepackage{amsfonts}
\usepackage{textcomp}
\usepackage{bbm}
\usepackage{setspace}
\usepackage{dsfont}

\usepackage{longtable}

\newcommand{\beq}{\begin{eqnarray}}
\newcommand{\eeq}{\end{eqnarray}}

\newcommand{\bmp}{\noindent\begin{minipage}{16cm}}
\newcommand{\emp}{\end{minipage}\vskip 7mm} 

\newcommand{\slashed}[1]{{#1}\hspace{-2mm}/}

\def\drawbox#1#2{\hrule height#2pt
        \hbox{\vrule width#2pt height#1pt \kern#1pt
              \vrule width#2pt}
              \hrule height#2pt}

\def\Asym#1#2{\vcenter{\vbox{\drawbox{#1}{#2}
              \kern-#2pt 
              \drawbox{#1}{#2}}}}

\def\simge{\mathrel{%
   \rlap{\raise 0.511ex \hbox{$>$}}{\lower 0.511ex \hbox{$\sim$}}}}

\def\simle{\mathrel{
   \rlap{\raise 0.511ex \hbox{$<$}}{\lower 0.511ex \hbox{$\sim$}}}}

\def\s#1{\setbox0=\hbox{$#1$}%
\rlap{\ifdim\wd0>.7em\kern.22\wd0\else\kern.1\wd0\fi /}#1}

\newcommand{\rem}[1]{}
\begin{document}

\begin{titlepage}
\title{\vspace*{-2.0cm}
\hfill {\small ICCUB-13-219, UB-ECM-PF-91-13}\\[3mm]
\bf\Large New Production Mechanism for keV Sterile Neutrino Dark Matter by Decays of Frozen-In Scalars
\\[1mm] }

\author{
Alexander Merle$^{a}$\thanks{email: \tt a.merle@soton.ac.uk},~~
Viviana Niro$^b$\thanks{email: \tt niro@ecm.ub.edu},~~~and~~
Daniel Schmidt$^c$\thanks{email: \tt Daniel.Schmidt@mpi-hd.mpg.de}
\\ \\
$^a${\normalsize \it Physics and Astronomy, University of Southampton,}\\
{\normalsize \it Highfield, Southampton, SO17 1BJ, United Kingdom}\\
\\
$^b${\normalsize \it Departament d'Estructura i Constituents de la Mat\`eria and Institut de Ci\`encies del Cosmos, }\\
{\normalsize \it Universitat de Barcelona, Diagonal 647, E-08028 Barcelona, Spain}\\
\\
$^c${\normalsize \it Max-Planck-Institut f\"ur Kernphysik, Saupfercheckweg 1,}\\
{\normalsize \it 69117 Heidelberg, Germany}
}
\date{\today}
\maketitle
\thispagestyle{empty}

\begin{abstract}
\noindent
We propose a new production mechanism for keV sterile neutrino Dark Matter. In our setting, we assume the existence of a scalar singlet particle which never entered thermal equilibrium in the early Universe, since it only couples to the Standard Model fields by a really small Higgs portal interaction. For suitable values of this coupling, the scalar can undergo the so-called freeze-in process, and in this way be efficiently produced in the early Universe. These scalars can then decay into keV sterile neutrinos and produce the correct Dark Matter abundance. While similar settings in which the scalar does enter thermal equilibrium and then freezes out have been studied previously, the mechanism proposed here is new and represents a versatile extension of the known case. We perform a detailed numerical calculation of the DM production using a set of coupled Boltzmann equations, and we illustrate the successful regions in the parameter space. Our production mechanism notably can even work in models where active-sterile mixing is completely absent.
\end{abstract}

\end{titlepage}

\section{\label{sec:intro}Introduction}

Our picture of the whole Universe has been strengthened first by the analysis of the WMAP 9-year data set~\cite{Hinshaw:2012fq} in combination with the data from the ground based telescopes SPT~\cite{Hou:2012xq} and ACT~\cite{Sievers:2013wk}, which had been supplemented in March 2013 by the release of the long awaited data obtained by the Planck satellite~\cite{Ade:2013lta}. Still we are puzzled by the ingredients of our Universe, one of the biggest mysteries being the identity of the so-called Dark Matter (DM). Even if the $\Lambda$CDM model, involving a cosmological constant $\Lambda$ and cold, i.e.\ non-relativistic, DM (CDM), provides a very good fit to the data~\cite{Ade:2013lta}, the intermediate case of warm Dark Matter (WDM) is still a valid possibility~\cite{Bode:2000gq,Hansen:2001zv,Boyarsky:2008xj,Lovell:2011rd,Boyanovsky:2010pw,Boyanovsky:2010sv,Boyanovsky:2010sv,VillaescusaNavarro:2010qy,Destri:2012yn,Destri:2013pt}. However, hot (i.e., highly relativistic) DM is clearly excluded by structure formation arguments~\cite{Abazajian:2004zh,dePutter:2012sh}, 

One particularly interesting candidate particle which in most settings turns out to be WDM would be a sterile [i.e., mainly a Standard Model (SM) singlet] neutrino with a mass of a few keV. If such a particle exists, in addition to two heavier (i.e., GeV) neutrinos which are nearly degenerate in mass, the resulting setting, called the $\nu$MSM~\cite{Asaka:2005an}, can indeed simultaneously accommodate for neutrino masses, for DM, and for the baryon asymmetry of the Universe~\cite{Canetti:2012vf,Canetti:2012kh}. However, while the $\nu$MSM can successfully \emph{accommodate} for such a peculiar set of sterile neutrinos, it does unfortunately not yield an \emph{explanation} for the required mass pattern. This fact has triggered the construction of a variety of models in the recent years, which try to give such an explanation. The ideas used to obtain light sterile neutrinos thereby range from the application of the Froggatt-Nielsen mechanism~\cite{Merle:2011yv,Barry:2011fp,Barry:2011wb}, over flavour symmetries~\cite{Shaposhnikov:2006nn,Lindner:2010wr,Araki:2011zg}, extra dimensions~\cite{Kusenko:2010ik,Adulpravitchai:2011rq,Takahashi:2013eva}, extensions of the seesaw mechanism~\cite{Barry:2011wb,Zhang:2011vh,Heeck:2012bz,Cogollo:2009yi,Dias:2005yh,Dias:2010vt,Babu:2004mj,Dev:2012bd}, composite neutrinos~\cite{Grossman:2010iq,Robinson:2012wu}, global symmetries~\cite{Allison:2012qn,Sayre:2005yh}, loop suppressions~\cite{Ma:2009gu}, to gravitational effects~\cite{Mavromatos:2012cc} -- see Ref.~\cite{Merle:2013gea} for a recent review. Even lighter sterile neutrinos have also attracted considerable interest (see, e.g., Refs.~\cite{Palazzo:2013me,Abazajian:2012ys} for two up-to-date reviews), but in general right-handed neutrinos have applications at various scales~\cite{Drewes:2013gca}.

A frequent ``problem'' with non-standard DM candidates such as keV sterile neutrinos is that they cannot be produced easily via the generic process of thermal freeze-out. This is simple to understand, since this mechanism requires particles to be in thermal equilibrium with the plasma in the early Universe, which does not work for \emph{sterile} neutrinos as their interactions are too weak. Nevertheless, sterile neutrinos will in general have slight admixtures to active neutrinos. Thus, they can be produced from time to time from the thermal plasma even though they never entered thermal equilibrium. For the case of keV sterile neutrinos, this simple scenario is called the Dodelson-Widrow (DW) mechanism~\cite{Dodelson:1993je} and it is nowadays known to be excluded by observations, in case that no primordial lepton asymmetry is present in the early Universe~\cite{Canetti:2012vf,Canetti:2012kh}. Indeed, a large enough primordial lepton asymmetry could lead to a resonant transition --the so-called Shi-Fuller mechanism~\cite{Shi:1998km}-- producing a considerable amount of sterile neutrinos with a cooler non-thermal spectrum, in addition to the ones produced by the DW mechanism. In this way, some bounds could be evaded. On the other hand, in frameworks where the SM gauge group is extended, the sterile neutrinos could be charged non-trivially under the full gauge group and be sterile only with respect to SM interactions. In this case, although this is not compulsory~\cite{Khalil:2008kp}, thermal production of keV neutrinos could be revived~\cite{Bezrukov:2009th,Nemevsek:2012cd}. However, this mechanism would generically produce too much DM and by this overclose the Universe, thus requiring some dilution by the production of additional entropy~\cite{Scherrer:1984fd}. Moreover, it could get into trouble with bounds from Big Bang nucleosynthesis~\cite{King:2012wg}.

Probably the most versatile production mechanism from a particle physics point of view is the non-thermal production of keV sterile neutrinos by the decays of particles~\cite{Kaplinghat:2005sy,Lin:2000qq,Hisano:2000dz,Kitano:2005ge,DiBari:2013dna}, in particular of singlet scalars. Examples of this production mechanism exists for the scalar being an inflaton~\cite{Shaposhnikov:2006xi,Bezrukov:2009yw} or a more general equilibrated scalar singlet particle~\cite{Kusenko:2006rh,Petraki:2007gq}. This case is particularly interesting because it tends to lead to smaller bounds on the mass of the keV neutrino, a desirable feature, since keV-neutrinos with too large masses could be in danger with X-ray bound. For a recent collections of observational bounds from the non-observation of the decay into a light neutrino and a photon, $N_1 \to \nu \gamma$, see Refs.~\cite{Canetti:2012vf,Canetti:2012kh,Merle:2013ibc} and references therein.\footnote{This bound only applies if active-sterile mixing exists in the first place. This is not necessarily the case in all settings, e.g., the sterile neutrinos could be odd under a $\mathds{Z}_2$ symmetry forbidding the decay into a light neutrino and a photon, see Refs.~\cite{Allison:2012qn,Sierra:2008wj,Gelmini:2009xd,Ma:2012if}.}

The aim of this paper is to study a variant of the scalar decay production mechanism discussed in Refs.~\cite{Kusenko:2006rh,Petraki:2007gq}. The decisive point is that the scalar $\sigma$, which decays into the keV neutrinos, $\sigma \to N_1 N_1$, has to be efficiently produced in the early Universe, as otherwise it would not be abundant enough to yield a significant amount of DM. In Ref.~\cite{Petraki:2007gq}, this point has been studied in great detail for the two cases of early and late freeze-out of the scalar. However, there is an alternative way to produce the scalar particle from the thermal plasma, the so-called \emph{freeze-in}~\cite{Hall:2009bx}. Similar to the DW production of keV steriles, we assume the scalar to have only very feeble interactions with the thermal plasma, so that it can -- albeit being produced from time to time in the early Universe -- never enter thermal equilibrium. The interaction strength is typically ruled by the so-called \emph{Higgs portal}, which allows any scalar singlet field\footnote{As we will see later on, $\sigma$ denotes the physical component of the field $S$.} $S$ to appear in the Lagrangian together with the SM-like Higgs field $H$ in a term of the form $\lambda (H^\dagger H) S^2$. The strength $\lambda$ of this interaction is not very much constrained, as we will point out in a dedicated section, but depending on its value this parameter decides about the thermal history of the singlet scalar. For example, if $\lambda \gtrsim 10^{-6}$, the scalar will enter the thermal equilibrium~\cite{Petraki:2007gq}. If it is unstable but its lifetime is large enough, it could then freeze-out as thermal relic and afterwards decay to produce keV sterile neutrino DM. On the other hand, for smaller values, $\lambda \sim 10^{-10}$, the scalar could freeze-in instead and, provided that it is heavy enough and stable (or at least very long lived), itself be the DM in the Universe. This case has been studied for the scalar either being a generic heavy singlet~\cite{Yaguna:2011qn} or a light pseudo Nambu-Goldstone boson~\cite{Frigerio:2011in}. However, what has not been studied up to now in the literature is the combination of the two settings, namely the \emph{freeze-in} of an \emph{unstable} scalar $\sigma$ which subsequently produces keV sterile neutrinos via its decay. The current study will close this gap and show that such a scalar FIMP (Feebly Interacting Massive Particle~\cite{Hall:2009bx}) can also lead to an interesting and valid possibility to produce keV sterile neutrino DM.

The paper is structured as follows: We first give an illustrative description of the idea behind the mechanism in Sec.~\ref{sec:idea}. The more technical details, such as a description of the model setting and of the Boltzmann equations to solve, as well as a discussion of the relevant bounds are provided in Sec.~\ref{sec:details}. Our actual results are presented in Sec.~\ref{sec:results}, before concluding in Sec.~\ref{sec:conc}. The appendices provide further technical details, such as definitions of the effective degrees of freedom (Appendix~A), remarks on the use of modified Bessel functions (Appendix~B), as well as detailed analytical derivations of the Boltzmann equations (Appendix~C) and of the free-streaming horizon (Appendix~D).

\section{\label{sec:idea} The basic idea: keV neutrino production by the decays of scalar FIMPs}

Before entering the technical details, we would like to give an illustration of how the proposed mechanism works. The decisive point of the production of any particles through ``late'' decays of a metastable species is that the parent particle has to be produced in the first place. While this might seem as a disadvantage, since two production stages are needed, it can in many circumstances actually be advantageous, because different constraints may hold for the two particles involved. For example, the keV sterile neutrino cannot be produced from the thermal plasma only: in case it enters thermal equilibrium, it is typically overproduced since it is relativistic at freeze-out~\cite{Bezrukov:2009th,Nemevsek:2012cd}. If it does not enter thermal equilibrium and is only produced non-resonantly by small admixtures, its spectrum is too warm if the correct abundance is produced. The constraints from structure formation~\cite{Boyarsky:2008xj} can then only be realised for relative large keV sterile neutrino masses which are in conflict with the constraints from the non-observation of the decay of the keV neutrino into a light neutrino and a photon~\cite{Canetti:2012vf,Canetti:2012kh}. The singlet scalar, instead, can be produced via thermal freeze-out, and then by its decay lead to a suitable abundance of keV sterile neutrino DM, while at the same time being compatible with all bounds~\cite{Kusenko:2006rh,Petraki:2007gq}.

In this paper we pursue a different path to produce the singlet scalar $\sigma$: if the Higgs portal coupling is small enough, $\lambda \ll 10^{-6}$~\cite{Kusenko:2006rh,Petraki:2007gq}, the scalar particle never enters thermal equilibrium (due to its feeble interactions), but it can still be produced by the plasma. This opens up a new region in the parameter space where a non-negligible abundance can be produced, which actually \emph{increases} for increasing $\lambda$, contrary to what would happen in thermal freeze-out. This idea is not new, but it was recently summarised and systematised in Ref.~\cite{Hall:2009bx}, where also the term FIMP (feebly interacting massive particle) was introduced. Furthermore, a scalar that has been produced in this way had not been studied before for the case of keV sterile neutrinos production.

In our setup, the physical singlet scalar $\sigma$ is produced via freeze-in from the thermal plasma. Approximately, it will have a spectrum with thermal shape, but with an overall suppression factor. This scalar $\sigma$ will then fully decay into keV neutrinos $N_1$ via the reaction $\sigma \to N_1 N_1$. Note that, in principle, it also couples to the heavier sterile neutrinos $N_{2,3}$. We assume this decay to be kinematically forbidden, since we consider $M_{2,3} \gg m_\sigma$. Thus, the decisive decay mode is $\sigma \to N_1 N_1$. On the other hand, it might be possible to construct interesting scenarios with $m_\sigma / 2 < M_{2,3} < m_\sigma$, which could open up channels like $\sigma \to N_1 N_{2,3}$. For simplicity, we will discard these possibilities and always assume $M_1 \ll m_\sigma \ll M_{2,3}$, keeping in mind that there are several models and mechanisms which can indeed generate such a mass pattern for Majorana sterile neutrinos~\cite{Merle:2011yv,Barry:2011fp,Barry:2011wb,Shaposhnikov:2006nn,Lindner:2010wr,Araki:2011zg,Kusenko:2010ik,Adulpravitchai:2011rq,Takahashi:2013eva,Zhang:2011vh,Heeck:2012bz,Cogollo:2009yi,Dias:2005yh,Dias:2010vt,Babu:2004mj,Dev:2012bd,Allison:2012qn,Sayre:2005yh,Ma:2009gu,Mavromatos:2012cc,Merle:2013gea}.

\begin{figure}[t]
\centering
\includegraphics[scale=0.8]{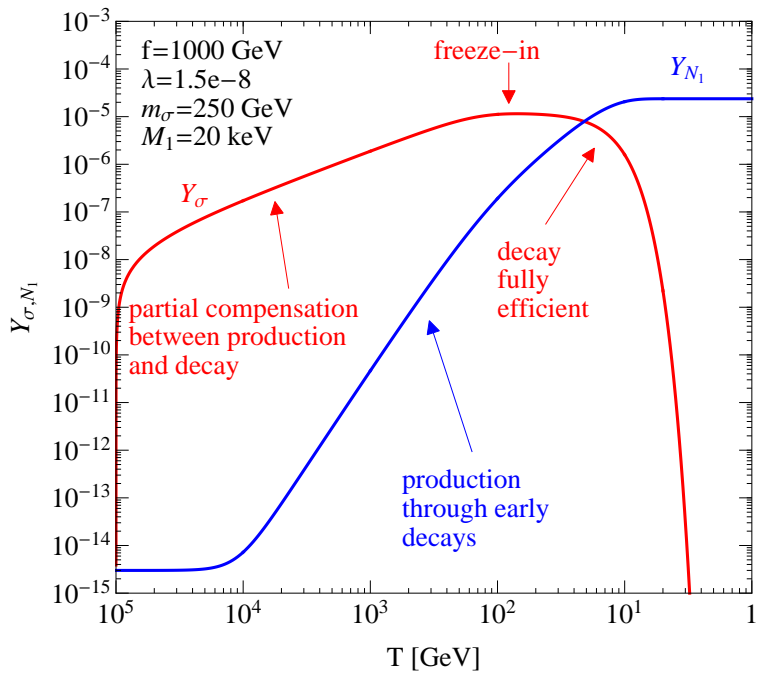}
\caption{\label{fig:Y_T}Example variation of yields $Y_{N_1}$ and $Y_\sigma$ as a function of the temperature $T$, cf.\ Sec.~\ref{sec:model} for details. As can be seen from the figure, a significant abundance of $\sigma$ gradually builds up due to freeze-in, before the decays $\sigma \to N_1 N_1$ set in and, at the same time, a significant amount of keV sterile neutrinos $N_1$ is produced.}
\end{figure}

An example evolution of the yields $Y$ of $\sigma$ and $N_1$ with decreasing temperature $T$ is displayed in Fig.~\ref{fig:Y_T}. As can be seen, we start essentially with a zero abundance of both particles (the precise value of the initial abundance plays no role as long as it is negligibly small), but with decreasing temperature the abundance of $\sigma$ increases before reaching a plateau at the freeze-in temperature $T \sim m_\sigma$. However, this abundance decreases again later, due to the decays $\sigma \to N_1 N_1$. Since every scalar $\sigma$ decays into exactly two $N_1$'s, this implies $Y_{N_1} (\text{late times}) = 2 Y_\sigma (\text{early times})$ as long as no $N_1$'s are produced from other sources. If the $N_1$'s are fully non-relativistic at late times, this also implies the relation $\Omega_{N_1} h^2 = 2\cdot \frac{M_1}{m_\sigma} \Omega_\sigma h^2$ between the final abundances, which makes it evident that this mechanism is useful to correct an overabundance of $\sigma$ by a suitable mass ratio $M_1/m_\sigma$. However, since the $N_1$'s can also be semi-relativistic (``warm''), at least for times close to their production, the above relation could receive a correction factor in case the yield $Y_{N_1}$ is not evaluated at a late enough time. Nevertheless, as an estimate, the above formula can be applied.

Finally, note that the assumption that the keV neutrinos $N_1$ are produced \emph{exclusively} by the described scalar decays does not always need to be true. In particular, in a setting where there is a non-negligible active-sterile mixing between $N_1$ and the light neutrinos $\nu_i$, a certain contribution to the abundance of $N_1$'s produced by the DW mechanism is unavoidable. We will take this contribution into account by estimating the maximal amount of keV neutrinos which can be produced by DW, without violating the X-ray bound or overproducing the DM. However, we would like to stress that our production mechanism \emph{does not need} active-sterile mixing. While such a mixing may or may not be desirable from a phenomenological perspective, there are settings known in which it is exactly zero~\cite{Allison:2012qn,Sierra:2008wj,Gelmini:2009xd,Ma:2012if}. In such a scenario the production of keV neutrinos by a combination of the standard DW and SF mechanisms would fail, while our mechanism (as well as the version where $\sigma$ does enter thermal equilibrium) could still be valid.

After having discussed the general idea behind our proposal, we will now present the more technical aspects of our work.


\section{\label{sec:details} Details of our analysis}

\subsection{\label{sec:model} The Model}

The particle content of the SM is extended by three right-handed sterile neutrinos $N_a$ $(a=1,2,3)$ and one real scalar singlet $S$~\cite{Petraki:2007gq}. The Lagrangian is
\begin{eqnarray}
 \cal L  &=&  {\cal L_{\rm SM}} + \left[ i \overline{N_a} \slashed{\partial} N_a + \frac{1}{2} (\partial_\mu S) (\partial^\mu S) - \frac{y_a}{2} S \; \overline{N_a^c} N_a \; + h.c.\, \right] - V_{\rm scalar} + \mathcal{L}_\nu \,,
 \label{eq:Lagrangian}
\end{eqnarray}
which consists of the SM, kinetic terms of the sterile neutrinos $N_a$, Yukawa interactions $f_a$ of the singlet $S$ with $N_a$, and a scalar potential $V_{\rm scalar}$. Finally, $\mathcal{L}_\nu$ is the part of the Lagrangian giving mass to the light neutrinos. In the simplest setting, we would have $\mathcal{L}_\nu = - y_D^{\alpha a} \overline{L_\alpha} \tilde H N_a + h.c.$ (where $\tilde H = i \sigma_2 H^*$). Then, a type~I seesaw mechanism~\cite{Minkowski:1977sc,Yanagida:1979as,GellMann:1980vs,Glashow:1979nm,Mohapatra:1979ia} could be at work using the right-handed Majorana masses for $N_a$ arising from a VEV $f = \langle S \rangle$, at least if the Yukawa couplings respect the observational X-ray bound~\cite{Merle:2012xq}. Alternatively, there could exist, e.g., more complicated seesaw-type mechanisms or radiative light neutrino mass generation~\cite{Ma:2006km,Zee:1985rj,Babu:1988ki,Krauss:2002px,Aoki:2008av}. Since we do not rely on a specific mechanism, we will leave the mass generation of light neutrinos unspecified. Any realistic setting must provide a way to generate a viable light neutrino mass and mixing pattern, but the details do not play a decisive role in our production mechanism.

We restrict our considerations to a potential $V_{\rm scalar}$ which only depends on the absolute value of the SM Higgs field $H$ and on even powers of the real scalar singlet $S$. Such a potential results from a global symmetry, e.g., lepton number, and does not impose a severe restriction. Assuming a global $\mathds{Z}_4 = \{ \pm 1, \pm i \}$ symmetry,\footnote{Note that this discrete $\mathds{Z}_4$ symmetry might potentially be problematic, since its breaking by a non-zero VEV $f = \langle S \rangle$ could lead to so-called~\emph{domain walls}~\cite{Zeldovich:1974uw}, which would considerably alter the history of the Universe but are not observed. There are arguments for how this problem could be evaded, see e.g.\ Refs.~\cite{Dvali:1995cc,Dvali:1996zr,Larsson:1996sp}. We will not enter this discussion here and simply assume that this problem is solved in a model containing the framework presented here. Nevertheless, we would like to point out that the most obvious solution of taking $S$ to be complex and promote the symmetry to a global $U(1)$ rotation, for which no domain walls would appear, is \emph{not} a straightforward solution to pursue. In that case, our production mechanism would suffer considerably from the existence of a Goldstone boson~\cite{Goldstone:1961eq} (more precisely a singlet Majoron~\cite{Chikashige:1980ui}) which would also couple to $N_1$ and considerably modify the DM production. In general, there can be a non-trivial interplay between the abundances of the different scalar fields in the early Universe, which makes the model with a complex scalar $S$ considerably different from the freeze-in of a real scalar, the latter case being addressed in this paper.} such that $S \to -S$ and $N_k \to i N_k$ (while all other fields transform trivially), the most general potential is:
\begin{equation}
 V_{\rm scalar} = -\mu_H^2 H^\dagger H - \frac{1}{2}\mu_S^2 S^2 + \lambda_H (H^\dagger H)^2 + \frac{1}{4} \lambda_S S^4 +2 \lambda (H^\dagger H)S^2.
 \label{eq:potential}
\end{equation}
The $SU(2)$ Higgs doublet $H\sim (\underline{\bf 2}, +1)$ and the scalar singlet $S~(\underline{\bf 1}, 0)$ are parametrized as
\begin{equation}
 H = \begin{pmatrix}
 h^+\\
 \frac{1}{\sqrt{2}} (v + \tilde{h} e^{i\rho})
 \end{pmatrix} \rightarrow \begin{pmatrix}
 0\\
 \frac{1}{\sqrt{2}} (v + \tilde{h})
 \end{pmatrix}\ \ \ {\rm and}\ \ \ S =f + \tilde{\sigma}.
 \label{eq:scalars}
\end{equation}
Note that the Goldstone bosons $h^\pm$ are eaten by $W^\pm$ to make them massive, similar to neutral boson $\rho$ being eaten by the $Z^0$. All other components are physical: $\tilde{h}$ is the SM-like Higgs and $\tilde{\sigma}$ is a physical singlet scalar. The VEVs are given by $\langle H\rangle=\frac{1}{\sqrt{2}}v$, where $v=246$~GeV (in our convention), and $\langle S \rangle = f$. Note that $f$ could potentially be large.

Inserting the VEVs, $H^\dagger H \rightarrow v^2/2$ and $S^2\rightarrow f^2$, and differentiating the potential with respect to $v^2$ and $f^2$, respectively, gives the minimum conditions
\begin{equation}
\left\{
\begin{matrix}
 \mu_H^2 = \lambda_H v^2 + 2\lambda f^2,\\
 \mu_S^2 = \lambda_S f^2 +2\lambda v^2.
\end{matrix}
\right.
 \label{eq:min-conds}
\end{equation}
The \emph{Higgs portal} coupling $\lambda$ results in mixing of the physical scalar fields. Concentrating on the potential terms which are proportional to $\tilde{\sigma}^2$, $\tilde{h}^2$, and $\tilde{\sigma} \tilde{h}$, and inserting the minimum conditions, Eq.~\eqref{eq:min-conds}, the mass matrix in the interaction basis $(\tilde{h},\tilde{\sigma})^T$ reads:
\begin{eqnarray}\label{eq:mass1}
 \begin{pmatrix}
\lambda_H v^2 & 2\lambda v f\\
2\lambda v f & \lambda_S f^2
\end{pmatrix}\,.
\end{eqnarray}
In the basis $(h, \sigma)^T$ of mass eigenstates we have, in the limit of small $\lambda$,
\begin{eqnarray}\label{eq:mass2}
 \frac{1}{2}\,(h, \sigma)
 \begin{pmatrix}
 m_h^2&0\\
 0&m_\sigma^2
 \end{pmatrix}
 \begin{pmatrix} 
 h\\
 \sigma
 \end{pmatrix}\, , \ \ \ {\rm where} \ \ \
 \left.
 \begin{matrix}
 m_h^2\\ 
 m_\sigma^2
 \end{matrix}
 \right\} \simeq
 \left\{
 \begin{matrix}
 \lambda_H v^2\\
 \lambda_S f^2
 \end{matrix}
 \right\}
 \mp \frac{(2 \lambda f v)^2}{\lambda_S f^2 - \lambda_H v^2},
\end{eqnarray}
Interpreting the transition from the interaction to the mass basis as an abstract rotation,
\begin{eqnarray}
\begin{pmatrix}\tilde{h}\\ \tilde{\sigma}\end{pmatrix}=
\begin{pmatrix}
\cos\alpha&-\sin\alpha\\ 
\sin\alpha&\cos\alpha
\end{pmatrix}
\begin{pmatrix}
h\\\sigma
\end{pmatrix}\,,
\end{eqnarray}
such that Eqs.~\eqref{eq:mass1} and~\eqref{eq:mass2} yield:
\begin{eqnarray}\label{eq:lambdas}
 && \lambda_S = \frac{m_h^2\sin^2\alpha+m_\sigma^2\cos^2\alpha}{2f^2}\,,\ \ \ \lambda_H = \frac{m_h^2\cos^2\alpha+m_\sigma^2\sin^2\alpha}{2v^2}\,,\nonumber\\
 && \lambda = \frac{(m_h^2-m_\sigma^2)\cos\alpha\sin\alpha}{4fv}\,.
\end{eqnarray}
The independent parameters are the singlet mass $m_{\sigma}$, the Higgs portal $\lambda$, and the VEV $f$ of the singlet. Since the Higgs portal $\lambda$ is small, we can practically identify $h$ with $\tilde{h}$ and $\sigma$ with $\tilde{\sigma}$. We will use the notation $h$ and $\sigma$ in the following. 

In our numerics, we have fixed the SM Higgs mass to 125~GeV in accordance with the experimental results by the ATLAS~\cite{ATLAS:2012oga,Tafirout:2012gc,ATLAS:2012cpa,ATLAS:2012toa,ATLAS:2012soa,ATLAS:2012ad,ATLAS:2012ac,ATLAS:2012zna,Aad:2012tfa,Aad:2012an} and the CMS~\cite{Chatrchyan:2012jja,Singh:2012fa,CMS:2012zwa,CMS:2012rwa,CMS:2012zta,Chatrchyan:2012qr,Chatrchyan:2012ww,Lai:2013yaa,Chatrchyan:2013yea} collaborations. In addition we assume $m_\sigma > m_h$ for definiteness and $m_\sigma < f$ in order to avoid being in danger to enter a non-perturbative regime, i.e., we vary $m_\sigma$ between the upper 1$\sigma$ limit of $m_h < 126.4$~GeV~\cite{Higgs-Talk} and $f$.

\subsection{\label{sec:relic_density} Relic density}

The relic density of our DM candidate particle $N_1$ is produced by the decays of a frozen-in real scalar singlet particle $\sigma$. The Boltzmann equations for the annihilation and the decay processes are given in Eqs.~\eqref{Boltzmann:n_A} and~\eqref{Boltzmann:n_D}, respectively. To calculate the relic density of $N_1$, we have to solve a system of coupled equations describing simultaneously the annihilation and decay processes as it is done in, e.g, Ref.~\cite{Adulpravitchai:2011ei}.

We have numerically solved the following two coupled Boltzmann equations: 
\begin{eqnarray}
\frac{d}{dT}Y_\sigma &=& \frac{d}{dT}Y_\sigma^{\mathcal{A}} + \frac{d}{dT}Y_\sigma^{\mathcal{D}}\,, \label{eq:Boltzmann_1}\\
\frac{d}{dT}Y_{N_1} &=& \frac{d}{dT}Y_{N_1}^{\mathcal{D}}\,, \label{eq:Boltzmann_2}
\end{eqnarray} 
with
\begin{eqnarray}
&& \frac{d}{dT}Y_\sigma^{\mathcal{A}} = -\sqrt{\frac{\pi}{45G_N}}\sqrt{g_\ast}\langle\sigma_{\rm ann} v\rangle\,Y_{\sigma,\rm eq}^2\,,\ \ \ 
\frac{d}{dT}Y_\sigma^{\mathcal{D}} = -\frac{1}{2}\frac{d}{dT}Y_{N_1}^{\mathcal{D}}\,, \nonumber\\
&& \frac{d}{dT}Y_{N_1}^{\mathcal{D}} = -\sqrt{\frac{45}{\pi^3G_N}}\frac{1}{T^3}\frac{1}{\sqrt{g_{\rm eff}}} \langle\Gamma(\sigma \rightarrow N_1 N_1)\rangle\,Y_\sigma \,,
\end{eqnarray}
see Appendix~\ref{sec:A_C}, in particular Eqs.~\eqref{yield:A} and \eqref{yield:D}, for detailed information. The equilibrium yield is given by 
\begin{equation}
Y_{\sigma, {\rm eq}}=\frac{45 g_\sigma}{4 \pi^4} \frac{x^2 K_2(x)}{h_{\rm eff}(T)}\,,
\end{equation}
with $g_\sigma = 1$ being the spin degrees of freedom for the particle $\sigma$, $x\equiv\frac{m_\sigma}{T}$, and 
\begin{equation}
\sqrt{g_\ast}\equiv\frac{h_{\rm eff}}{\sqrt{g_{\rm eff}}} \left(1 + \frac{1}{3}\frac{T}{h_{\rm eff}}\frac{dh_{\rm eff}}{dT}\right)\,.
\end{equation}
For the definitions of $h_{\rm eff}$, $g_{\rm eff}$ and of the Bessel functions, see Appendices~A and~B. As already explained, the DM particle is the lightest sterile neutrino $N_1$ which is produced by the frozen-in real scalar singlet $\sigma$ due to out-of-equilibrium decays, $\sigma\rightarrow N_1N_1$. The thermally averaged cross section times relative velocity $\langle\sigma_{\rm ann} v\rangle$ for the real scalar singlet $\sigma$ is calculated numerically using the micrOMEGAs package~\cite{Belanger:2010gh}. $\langle\Gamma(\sigma \rightarrow N_1 N_1)\rangle$ is the thermally averaged decay rate for the decay $\sigma\rightarrow N_1 N_1$ and the analytically determined decay width in the rest frame of the decaying particle $\sigma$ is
\begin{equation}
 \Gamma (\sigma \rightarrow N_1 N_1) = \frac{y_1^2}{16\pi} m_\sigma \left[ 1-\frac{4 M_1^2}{m_\sigma^2} \right]\,.
 \label{eq:rate}
\end{equation}
See Eqs.~\eqref{eq:decay1} and~\eqref{eq:decay2} for the definition of $\langle\Gamma(\sigma \rightarrow N_1 N_1)\rangle$. Finally, the relic density can be obtained using the following formula: 
\begin{equation}
\Omega_{\rm DM} h^2 = 2.733 \times 10^8\,\frac{m_{\rm DM}}{\rm GeV}\,Y_0\,,
\end{equation}
with $Y_0=Y_{N_1}(T_0)$ being the yield of the DM particle at late times.

\subsection{\label{sec:horizon}Existing constraints on the free streaming horizon}

In order to determine whether the neutrinos generated act as CDM or WDM, one would actually need to determine the velocity profile and do a full simulation of the resulting structures in the Universe, see e.g.\ Ref.~\cite{Colombi:1995ze}. However, this is a big effort and far beyond the scope of this paper. Alternatively, one gets at least an indication by computing the so-called \emph{(co-moving) free-streaming horizon} $r_{\rm FS}$, which can be interpreted as the mean distance which the DM particles would travel if they were not bound by gravitation at some point. The free-streaming horizon is defined as~\cite{Boyarsky:2008xj}
\begin{equation}
 r_{\rm FS} = \int\limits_{t_{\rm in}}^{t_0} \frac{\langle v(t) \rangle}{a(t)} dt,
 \label{eq:FS-horizon}
\end{equation}
where $t_{\rm in}$ is the initial time at which the integration starts, $t_0$ is the current time, $v(t)$ is the mean velocity of the DM particles, and $a(t)$ is the scale factor. The free-streaming horizon is a co-moving quantity and, as we will see, one can define a free-streaming horizon of $0.1$~Mpc~\cite{Colin:2000dn}, which is about the size of a dwarf galaxy, as the separation between HDM ($\lambda_{\rm FS} > 0.1$~Mpc) and WDM ($\lambda_{\rm FS} < 0.1$~Mpc). In turn, free-streaming horizons which are considerably smaller typically correspond to CDM. Note that this is in some sense an artificial definition, as we will explain in detail in Sec.~\ref{sec:results}, but it nevertheless gives a good orientation in practice. As we will see, the condition $r_{\rm FS} < 0.1$~Mpc will lead to a \emph{lower} bound on the mass $M_1$ of the keV sterile neutrino. We will compute this bound by an approximate solution of the integral in Eq.~\eqref{eq:FS-horizon}.\footnote{Note that this is similar but not equivalent to the early freeze-out results from Ref.~\cite{Petraki:2007gq}. Taking the numerical calculation from that reference, we can indeed reproduce our results within a factor of $2$.}

Let us start with the integral boundaries. The production time of the DM particles can be approximated by $t_{\rm in} \equiv t_{\rm prod} + \tau$, where $t_{\rm prod}$ is the time of freeze-in (i.e., the time when the temperature equals the FIMP mass $m_\sigma$~\cite{Hall:2009bx}) and $\tau = 1/ \Gamma$ is the lifetime of the scalar particle $\sigma$, cf.\ Eq.~\eqref{eq:rate}. The scale factor $a(t)$ can be approximated as $a(t) \propto t^{1/2}$ [$a(t) \propto t^{2/3}$] for radiation [matter] dominance. Note that it is perfectly fine to neglect the vacuum-dominated part of the integral in Eq.~\eqref{eq:FS-horizon} and to assume matter-dominance until $t_0$, since very late times practically do not have any effect on the result~\cite{Boyarsky:2008xj}. This treatment is perfectly motivated and serves as an easy approximation. However, one still has to take into account the entropy dilution from the time of production, which happens at a very high temperature, to the current time. This amounts to a further factor of $\xi^{-1/3}$~\cite{Petraki:2007gq}, with an entropy dilution factor given by
\begin{equation}
 \xi = \frac{g_{\rm eff} ({\rm high}\ T)}{g_{\rm eff} (t_0)} \approx \frac{109.5}{3.36},
 \label{eq:dilution-factor}
\end{equation}
where we have taken both the real scalar $\sigma$ and the keV Majorana neutrino $N_1$ to contribute to radiation at high temperatures.\footnote{At this step, we disagree with Ref.~\cite{Petraki:2007gq}, where the number of degrees of freedom at a high temperature has been taken to be $110.5$. This corresponds to one Majorana neutrino and a \emph{complex} scalar~\cite{Vilja:1993uw}. However, the resulting numerical difference is tiny and would in no case affect the results significantly.} Since the scalar $\sigma$ has been produced in a significant amount at the time of its decay, this should not be a bad approximation, and for the same reason also a significant amount of $N_1$'s should be around.

The crucial question is how to determine the average velocity $\langle v(t) \rangle$. For simplicity, we assume an instantaneous transition between the highly relativistic and the fully non-relativistic regimes,
\begin{equation}
 \langle v(t) \rangle \simeq \left\{
 \begin{matrix}
 1 & & & \text{if $t < t_{\rm nr}$,} \hfill\\
 \frac{\langle p(t) \rangle}{M_1} & & & \text{if $t \geq t_{\rm nr}$,}
 \end{matrix}
 \right.
 \label{eq:FS_trans}
\end{equation}
where $t_{\rm nr}$ is the time when the particle becomes non-relativistic, defined by the equality between its average momentum and its mass, $\langle p(t) \rangle = M_1$. This average momentum can be extracted from the distribution function of the DM particles. For non-relativistic parent particles $\sigma$ (which is a good approximation~\cite{Hall:2009bx}), this distribution function is given by~\cite{Kaplinghat:2005sy,Strigari:2006jf,Aoyama:2011ba,Kamada:2013sh}
\begin{equation}
 f(p,t) = \frac{\beta}{p/T_{\rm DM}} \exp \left(- \frac{p^2}{T_{\rm DM}^2} \right),
 \label{eq:FSdist_1}
\end{equation}
where $\beta$ is a normalization factor that will turn out to be irrelevant for our purposes, $p$ is the co-moving momentum, and the DM temperature is defined as $T_{\rm DM} = T_{\rm DM}(t) = p_{\rm cm} \ a(t_d) / a(t)$. Here, $p_{\rm cm} = \frac{\sqrt{m_\sigma^2 - M_1^2}}{2} \simeq \frac{m_\sigma}{2}$ is the DM momentum in the center-of-mass frame and the decay time $t_d$ is defined as $H(t = t_d) = \frac{1}{2 t_{\rm in}}$~\cite{Kamada:2013sh}. Since the particle production happens during radiation dominance, we know that $H(t_d) = 1/(2 t_d)$ and can thus identify $t_d \equiv t_{\rm in}$.

Defining ``early'' and ``late'' production of the DM particles as $t_{\rm in} < t_{\rm eq}$ and $t_{\rm in} > t_{\rm eq}$, respectively, where $t_{\rm eq}$ is the time of matter-radiation equality, one can easily compute the free-streaming horizon by splitting the integral from Eq.~\eqref{eq:FS-horizon} into different pieces for radiation/matter dominance and highly relativistic/non-relativistic DM particles. The result is 
\begin{equation}
 r_{\rm FS} \simeq \left\{
 \begin{matrix}
 \frac{\sqrt{t_{\rm eq} t_{\rm nr}}}{a_{\rm eq}} \left[ 5 + \ln \left( \frac{t_{\rm eq}}{t_{\rm nr}} \right) \right]/\xi^{1/3} \hfill \hfill \hfill \hfill \hfill \hfill & & & \text{if $t_{\rm nr} < t_{\rm eq}$,} \hfill\\
 \left[ \frac{3 t_{\rm eq}^{2/3} t_{\rm nr}^{1/3}}{a_{\rm eq}} - \frac{t_{\rm eq}}{a_{\rm eq}} + \frac{\sqrt{\pi}}{2} \frac{m_\sigma/2}{M_1} \sqrt{\frac{t_{\rm in}}{t_{\rm eq}}} 
\frac{3\ t_{\rm eq}^{4/3} }{a_{\rm eq} t_{\rm nr}^{1/3}} \right]/\xi^{1/3} & & & \text{if $t_{\rm nr} > t_{\rm eq}$,}
 \end{matrix}
 \right.
 \label{eq:FS_hor}
\end{equation}
see Appendix~\ref{sec:A_D} for details and in particular Eqs.~\eqref{eq:FS-horizon_early} and \eqref{eq:FS-horizon_late}. Note that the two parts of Eq.~\eqref{eq:FS_hor} coincide for $t_{\rm nr} \to t_{\rm eq}$. Furthermore, as to be expected, $r_{\rm FS}$ always increases with increasing $t_{\rm in}$. We have used Eq.~\eqref{eq:FS_hor} to mark the excluded region of HDM ($r_{\rm FS} > 0.1$~Mpc) later on in Figs.~\ref{fig:Omega_lambda} and~\ref{fig:Omega_lambda_2}. We will furthermore indicate the CDM regions ($r_{\rm FS} < 0.01$~Mpc), which are not excluded and instead reveal that also a keV-mass particle can act as CDM, depending on the details of its production.

\subsection{\label{sec:collider}Collider bounds on the production of Dark Matter }

In colliders, a DM signal can be detected through monojet or cascade events. If the DM particle is stable, it does not decay inside the detector volume and thus leaves its track as missing energy, which can be reconstructed. Comparing simulated DM events with data analysis allows to constrain the DM interaction and its mass. Bounds exist on DM masses around $1$~GeV. Specific collider constraints on Majorana fermion DM can be found in \cite{Goodman:2010yf}; for Dirac fermion, complex scalar, and real scalar DM, see \cite{Goodman:2010ku}. Note that these constraints are not relevant in our case, since the DM is a sterile neutrino $N_1$ with a mass in the keV range, produced by the decay of a frozen-in scalar singlet FIMP.

The scalar singlet FIMP itself is produced via the Higgs portal. In Ref.~\cite{Kamenik:2012hn}, the allowed region for the Higgs portal coupling $\lambda$ and the mass of the scalar singlet is presented. From that reference it follows that the strongest upper bound on the Higgs portal is $\lambda<0.01$, but for scalar singlet masses $m_\sigma \gtrsim 60 \rm{~GeV}$ there is essentially no constraint. In the mechanism we propose, the Higgs portal coupling is of order $\lambda\sim\mathcal{O}(10^{-8})$, i.e., given the mass range of our scalar singlet and its feeble interactions the constraints in Ref.~\cite{Kamenik:2012hn} for the allowed $\lambda-m_S$ region are not relevant for us. For completeness, see also \cite{Kanemura:2010sh} for LHC sensitivities on the Higgs portal.

In addition, Ref.~\cite{Kamenik:2012hn} presents constraints on the invisible decay width of the SM Higgs doublet $H$; for $m_h = 125$~GeV, the invisible decay width has to be smaller than approximately $0.0025$~GeV. In our model, the SM Higgs doublet $H$ decays invisibly into the DM particle $N_1$ and an active neutrino at tree-level, if light neutrino masses are generated by type I seesaw or by any other mechanism which allows for a neutrino Yukawa coupling given by $\mathcal{L}_\nu=-y_D^{\alpha 1}\overline{L_\alpha} \tilde H N_1$. Since $\langle H\rangle$ is of the order $\mathcal{O}(100)\rm{~GeV}$, $M_1$ of order $\mathcal{O}(1-100)\rm{~keV}$, and the light neutrino masses in the sub-eV range, the Yukawa couplings $y_D^{\alpha 1}$ must be tiny such that the decay width of $H\rightarrow N_1\nu_\alpha$ is much smaller than $0.0025$~GeV.

To conclude, all existing collider bounds on production of DM are not relevant for our mechanism and do not constrain the parameters we are considering in our numerical analysis.

\section{\label{sec:results}Results}

\begin{figure}
\centering
\begin{tabular}{lr}
\includegraphics[scale=0.4]{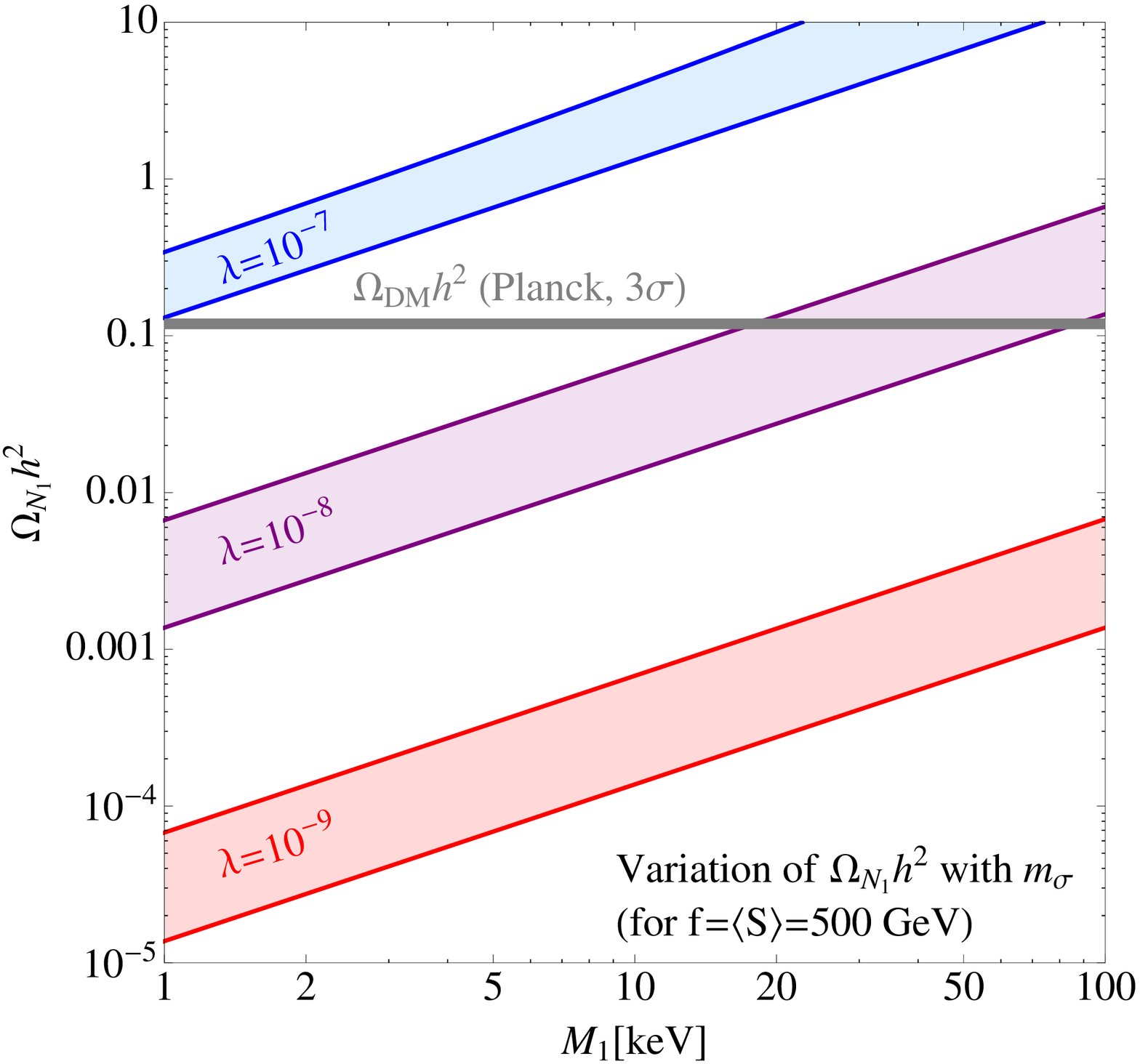} &
\includegraphics[scale=0.4]{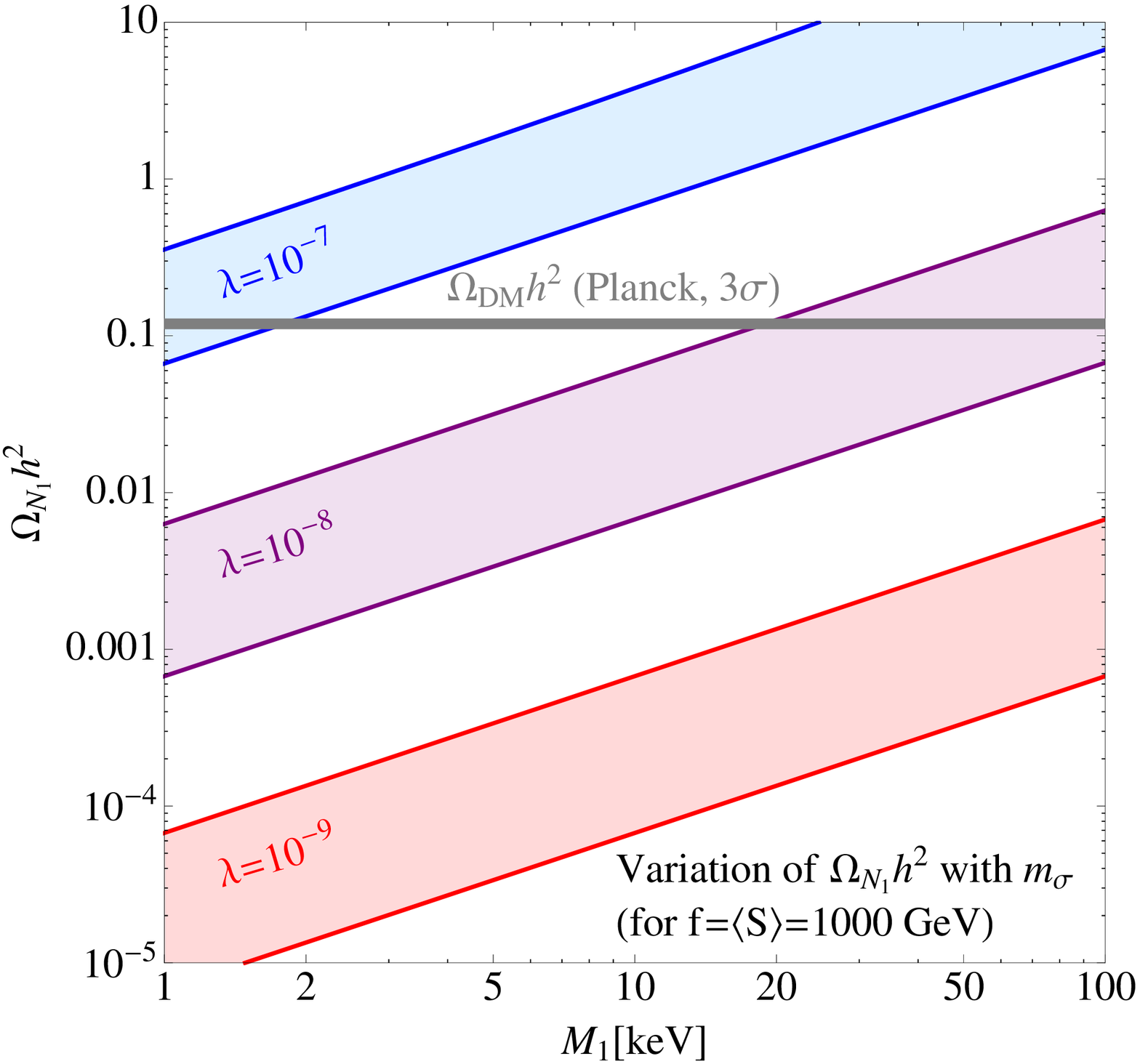}
\end{tabular}
\caption{\label{fig:Relic_lambda}
Relic density $\Omega_{N_1} h^2$ as a function of the sterile neutrino mass $M_1$, for Higgs portal coupling $\lambda = 10^{-7, 8, 9}$. In the left panel we show the results considering $f=500$~GeV, while in the right panel the results considering $f=1$~TeV.}
\end{figure}

We have numerically solved Eqs.~\eqref{eq:Boltzmann_1} and~\eqref{eq:Boltzmann_2} in order to determine the final abundance of keV sterile neutrinos $N_1$. First of all we scanned over a range of values for the Higgs portal coupling $\lambda$ in order to identify the successful region to obtain the correct relic abundance. The only requirement we impose on $\lambda$ is that $\lambda \lesssim 10^{-6}$ in order not to enter thermal equilibrium~\cite{Petraki:2007gq}. The result of this scan can be found in Fig.~\ref{fig:Relic_lambda}, where we plot the abundance regions for different values of the coupling, $\lambda = 10^{-7, 8, 9}$, as a function of the keV neutrino mass $M_1$. The broadening of the corresponding bands originates from the variation over the scalar mass $m_\sigma$. For definiteness, we assume that the singlet scalar mass is always larger than the SM-like Higgs mass $m_h \approx 125$~GeV (corresponding to the upper end of the bands in the plot). Furthermore, in order to avoid entering a potentially non-perturbative regime, we also assume that $m_\sigma < f$ (corresponding to the lower ends of the bands in the plot), where $f$ is the VEV of the singlet field $S$. In Fig.~\ref{fig:Relic_lambda}, we present the plots for the two example values $f = 500$~GeV and $f = 1$~TeV, which are perfectly compatible with all bounds. As can be seen from Fig.~\ref{fig:Relic_lambda}, the successful value of the Higgs portal coupling $\lambda$ should be around $10^{-8}$, more or less independently of the value of the VEV $f$. Accordingly, we will focus on the region $\lambda \approx 10^{-8}$ in what follows and investigate this region in greater detail in what concerns the relic abundance and in particular the experimental and observational bounds.

\begin{figure}
\centering
\begin{tabular}{lr}
\includegraphics[scale=0.45]{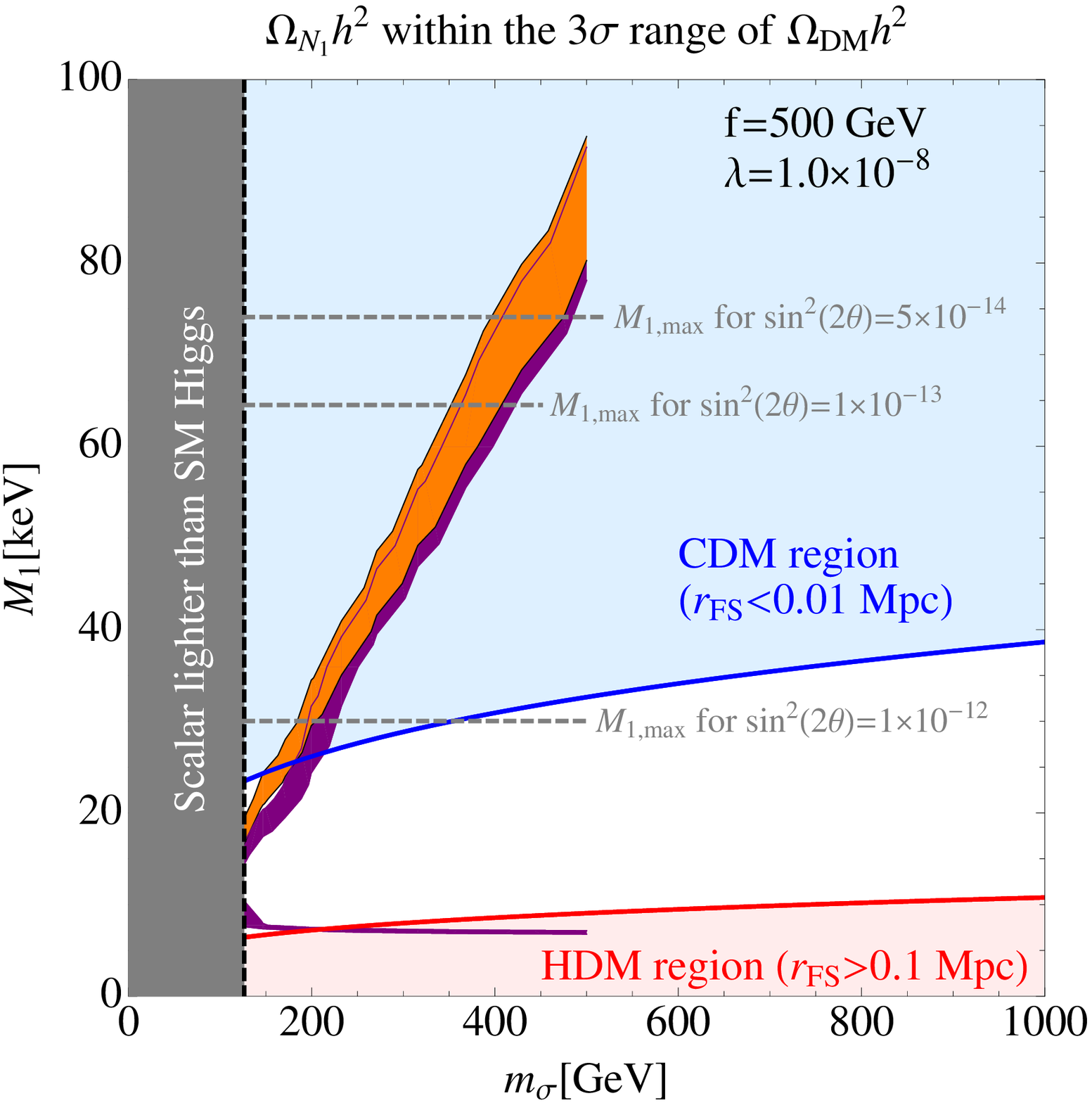} & \includegraphics[scale=0.45]{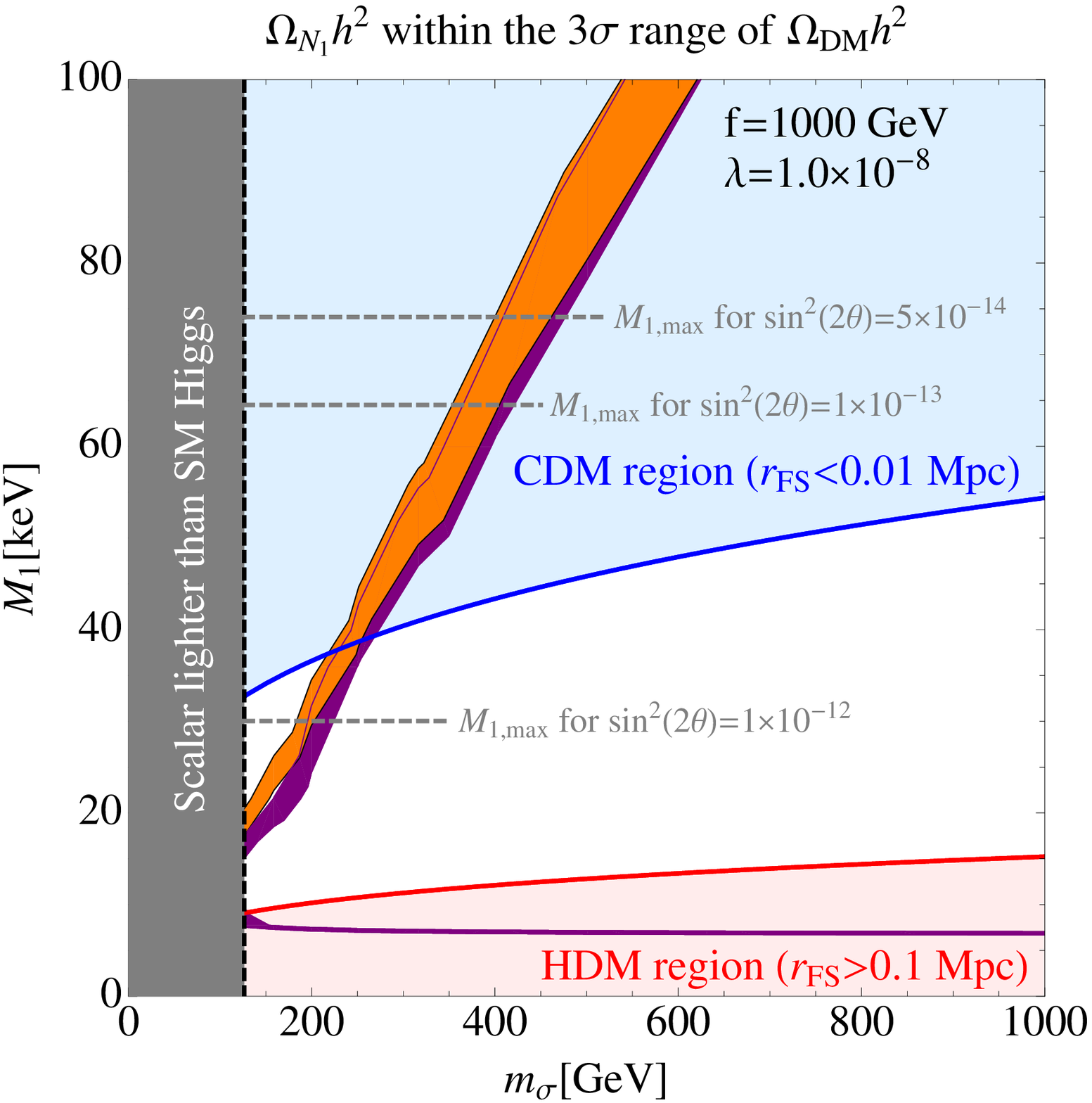}\\
\includegraphics[scale=0.45]{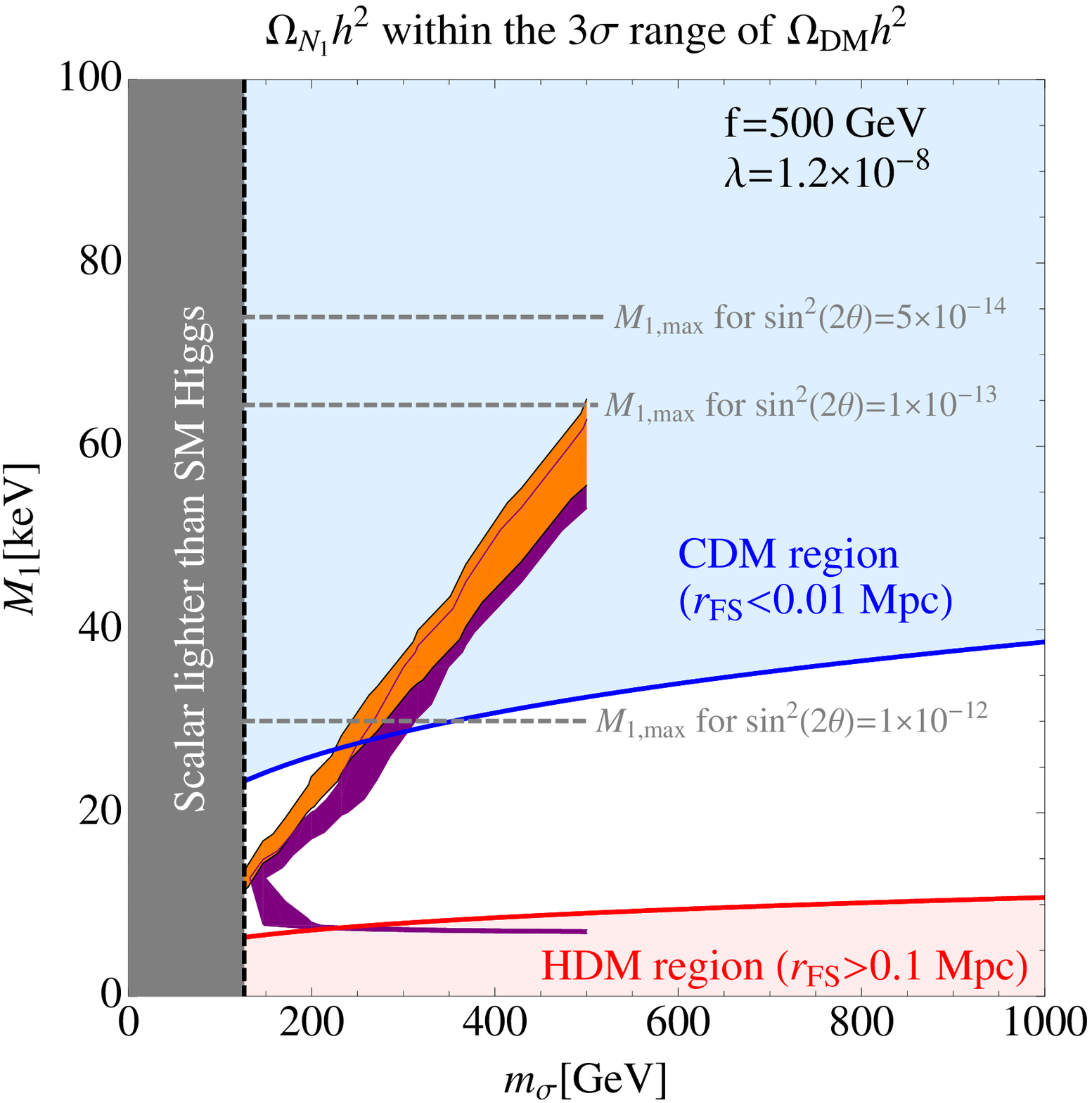} & \includegraphics[scale=0.45]{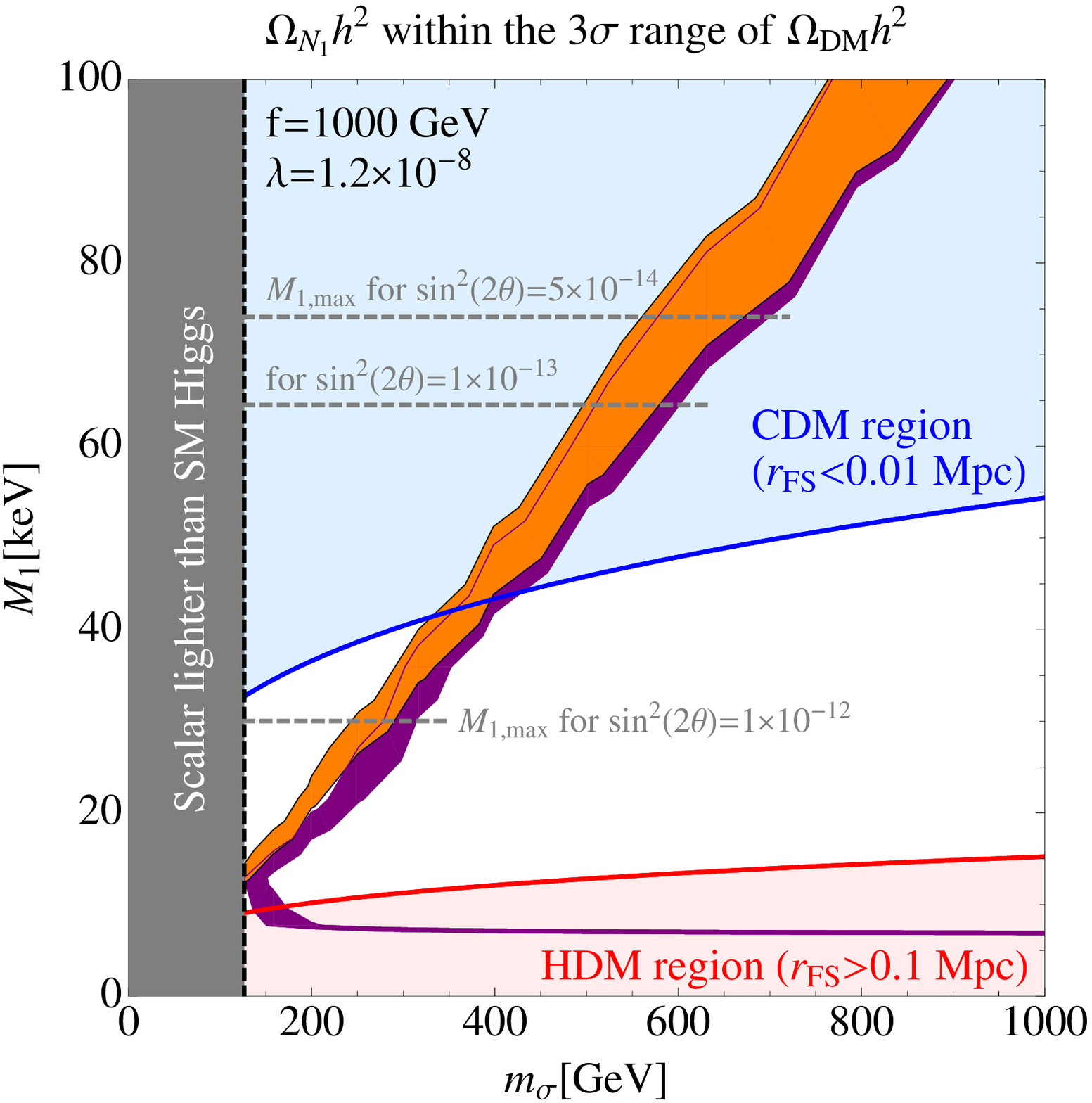}
\end{tabular}
\caption{\label{fig:Omega_lambda}We show the results considering $\lambda = 1.0\cdot 10^{-8},\ 1.2\cdot 10^{-8}$, as well as $f=500$~GeV and $f=1$~TeV. The orange (purple) bands represent the regions of the parameter space with a sterile neutrino relic abundance $\Omega_{N_1} h^2$ within the $3\sigma$ observed value, obtained only through the decay of a freeze-in scalar (considering also the DW mechanism), see text for more details. The red and blue areas denote the HDM and CDM regions, respectively.}
\end{figure}

\begin{figure}
\centering
\begin{tabular}{lr}
\includegraphics[scale=0.45]{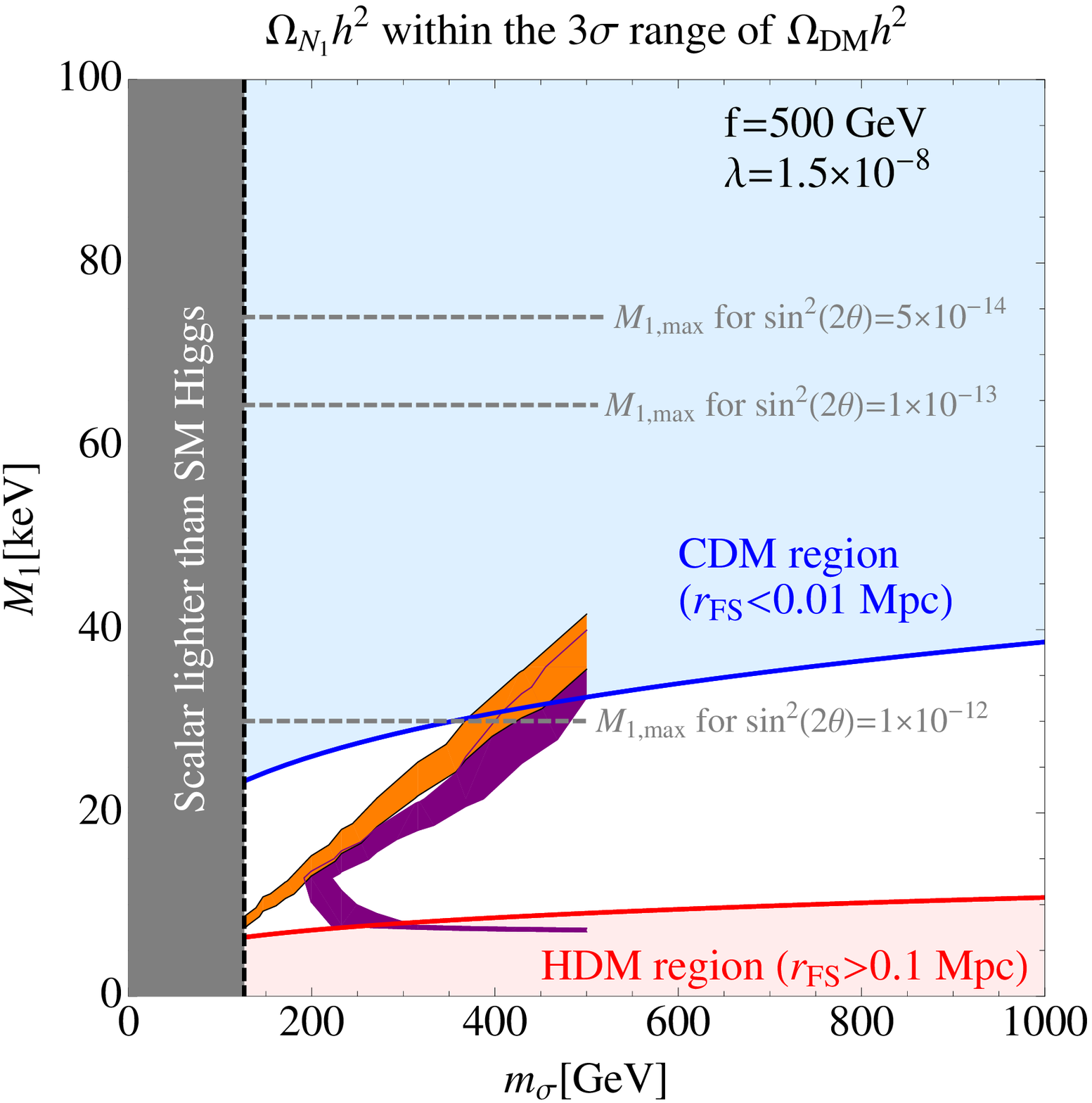} & \includegraphics[scale=0.45]{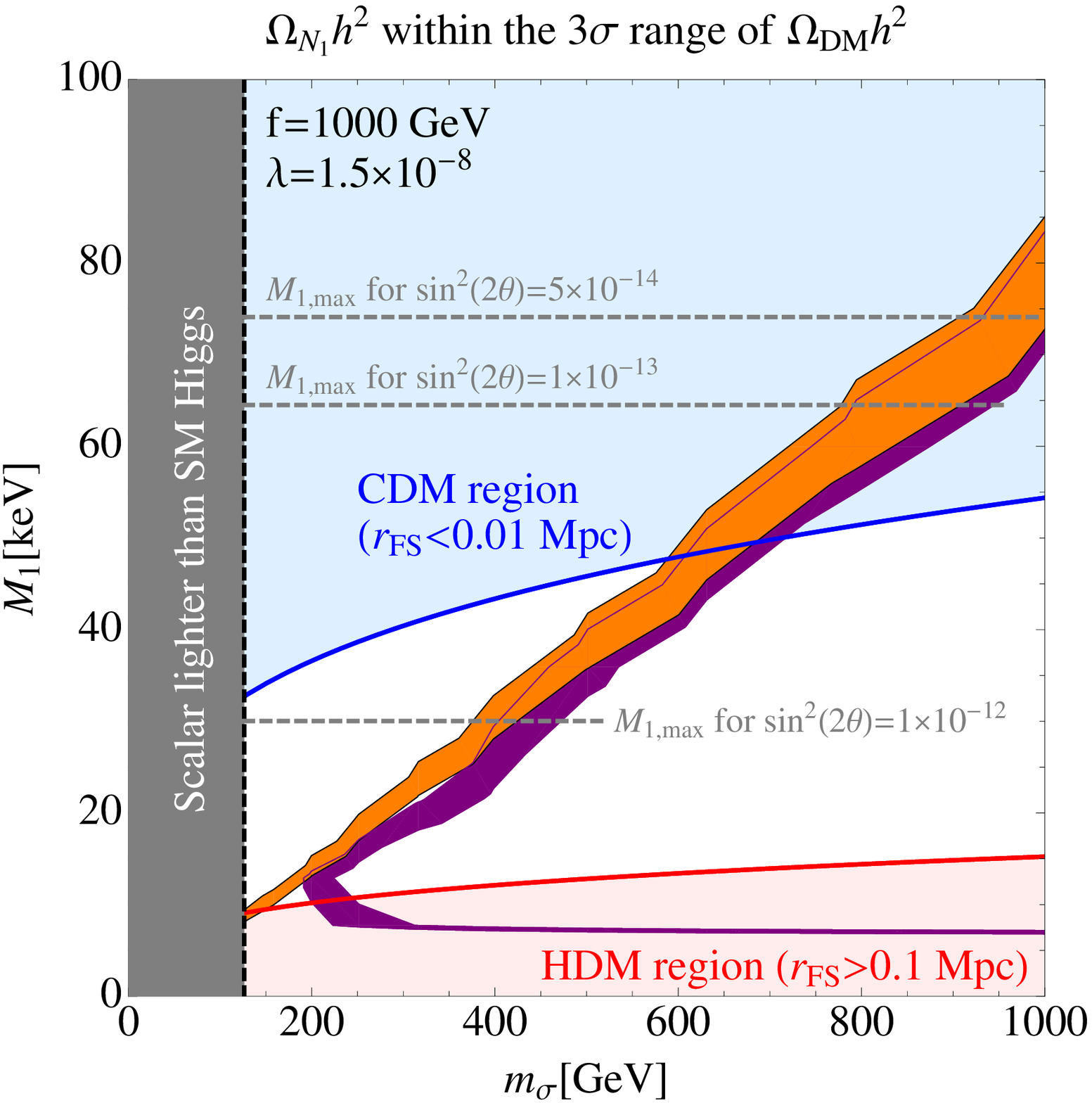}\\
\includegraphics[scale=0.45]{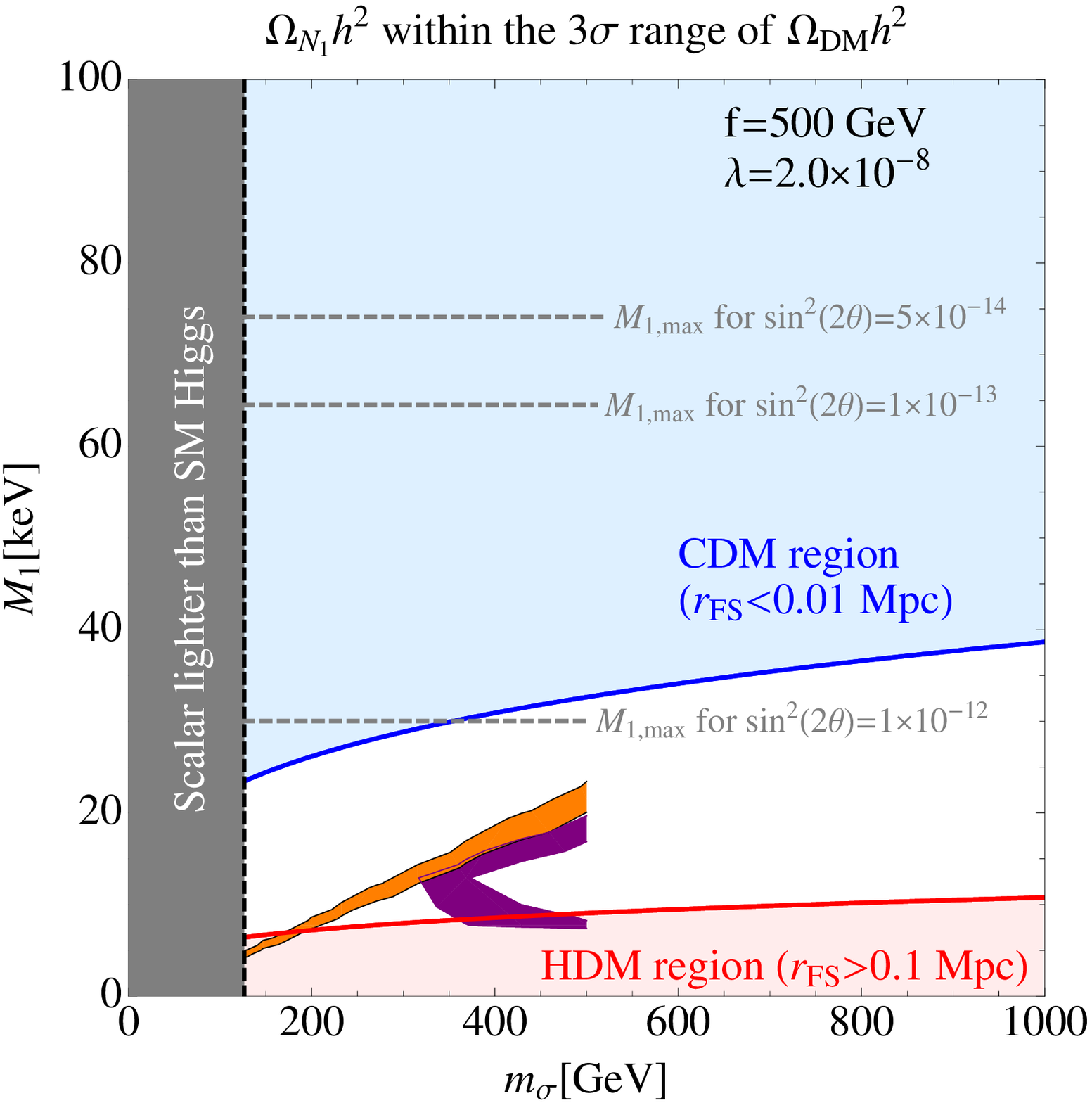} & \includegraphics[scale=0.45]{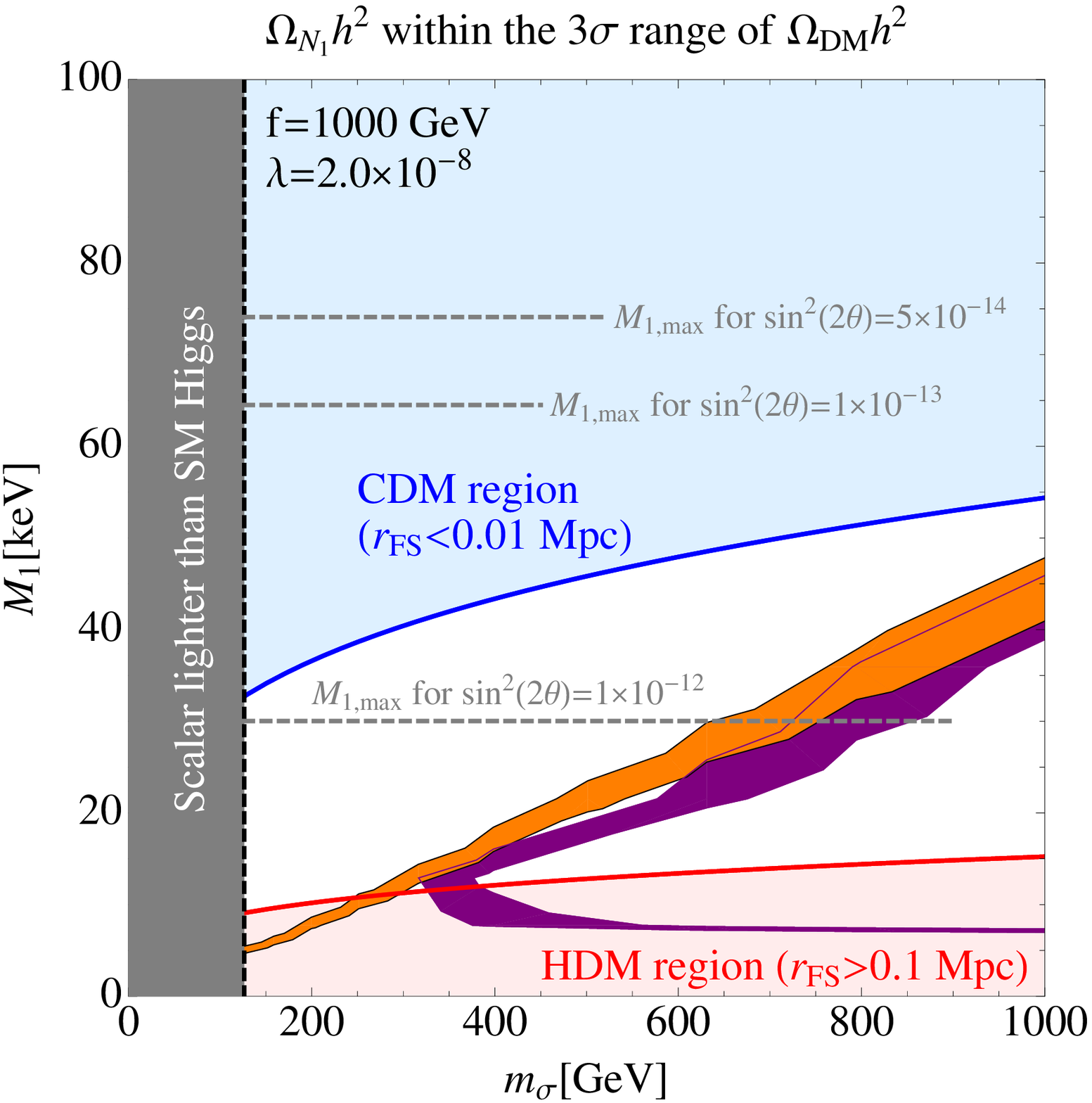}
\end{tabular}
\caption{\label{fig:Omega_lambda_2}Same ad Fig.~\ref{fig:Omega_lambda}, but for $\lambda = 1.5\cdot 10^{-8},\ 2.0\cdot 10^{-8}$.}
\end{figure}

A more detailed investigation of the successful regions in parameter space can be found in Figs.~\ref{fig:Omega_lambda} and~\ref{fig:Omega_lambda_2}, where we have indicated the region of the correct abundance (i.e., within the 3$\sigma$ ranges of the Planck data~\cite{Ade:2013lta}), as generated by scalar FIMP production only, by the orange band in the plot. The parameter values for the plots are chosen as $f \in \{ 500~{\rm GeV} , 1~{\rm TeV} \}$, with $\lambda \in \{ 1.0\cdot 10^{-8},\ 1.2\cdot 10^{-8} \}$ for Fig.~\ref{fig:Omega_lambda} and $\{1.5\cdot 10^{-8},\ 2.0\cdot 10^{-8} \}$ for Fig.~\ref{fig:Omega_lambda_2}. As can be seen from the plots, the iso-abundance lines reveal a more or less linear dependence of the keV sterile neutrinos mass $M_1$ on the scalar singlet mass $m_\sigma$. This feature can be understood easily by observing that the final DM energy density must be equal to the initial energy density in scalar $\sigma$ particles, which can at most be redshifted. Since this initial energy density is non-relativistic, it can be written as $\rho_\sigma = m_\sigma n_\sigma$, where $n_\sigma$ is the number density of $\sigma$-particles. Similarly, the energy density in $N_1$ can be computed by the non-relativistic expression for late times, cf.\ discussion in Sec.~\ref{sec:idea}, since in the successful regions in the parameter space the DM particles become non-relativistic within the age of the Universe.

In addition, we have indicated some important bounds. As explained, we have assumed that $m_\sigma > m_h$, and by gray rectangles we indicate the corresponding regions left of the upper 1$\sigma$ bound on $m_h$ of $126.4$~GeV~\cite{Higgs-Talk}. Furthermore, we know that HDM is excluded or, rather, bound to make up at most 1\% of the DM in the Universe~\cite{Abazajian:2004zh,dePutter:2012sh} by considerations of cosmological structure formation. A rough way to quantify when DM particles are HDM, WDM, or CDM is the co-moving free-streaming horizon $r_{\rm FS}$, cf.\ Sec.~\ref{sec:horizon}. Since it is a bit crude to attribute the property of a whole velocity spectrum of DM particles to one single number, it is to some extent a question of definition where to draw the lines between the three DM categories. A relatively common choice, which somewhat representatively reflects the use of the three terms in the literature~\cite{Sigurdson:2009uz,Das:2010ts} is to take the border between HDM and WDM at a free-streaming horizon of roughly $r_{\rm FS} = 0.1~{\rm Mpc}$, where larger values signal HDM which is forbidden. Note that this value is physically motivated due to the size of dwarf satellite galaxies being in that range. However, between CDM and WDM there is not a very well-defined boundary, since it is not easy to unambiguously define at which value of $r_{\rm FS}$ the structure formation on small scales starts to depart from the pure CDM case~\cite{Boyarsky:2008xj,Green:2005fa,Schneider:2013ria}. However, it is clear that the free-streaming horizon for CDM should be ``significantly smaller'' than the one for WDM. For definiteness, we have therefore decided to simply take a value that is by one order of magnitude smaller than the one for the HDM--WDM boundary. Keeping in mind that this distinction between WDM and CDM is also a matter of definition, the values of $r_{\rm FS}$ which we used are:
\begin{eqnarray}
 \textrm{Cold Dark Matter (CDM)}\ \ \ &:\Longleftrightarrow&\ \ \ r_{\rm FS} < 0.01~{\rm Mpc}, \nonumber \\
 \textrm{Warm Dark Matter (WDM)}\ \ \ &:\Longleftrightarrow&\ \ \ 0.01~{\rm Mpc} < r_{\rm FS} < 0.1~{\rm Mpc}, \nonumber \\
 \textrm{Hot Dark Matter (HDM)}\ \ \ &:\Longleftrightarrow&\ \ \ 0.1~{\rm Mpc} < r_{\rm FS}. \nonumber
\end{eqnarray}
Thus, in the plots displayed in Figs.~\ref{fig:Omega_lambda} and~\ref{fig:Omega_lambda_2}, the thick red line at the bottom of the plots marks the transition between WDM and HDM, and the light red region below this line is excluded by structure formation. The light blue region in the upper part of the plots, bounded by the thick blue line, corresponds to CDM and the white region marks the WDM sector. Here, it is worth to point out that in a considerable region of the parameter space, keV sterile neutrinos (with large enough masses) can be \emph{cold} DM (or, more precisely, indistinguishable from CDM according to our definition), in contrast to most of the scenarios for keV sterile neutrino DM. To some extent, this is a simple reflection of the fact that our DM production happens in the early Universe, but the more crucial point is that the mass ratio $m_\sigma / M_1$ happens to be in the correct region to allow for a sufficient cooling time. Figs.~\ref{fig:Omega_lambda} and~\ref{fig:Omega_lambda_2} reveal that the region of the correct DM relic abundance lies in the cold or warm DM parameter space, depending on the specific value of $\lambda$, respectively.


We also have the possibility to produce part of the DM in keV sterile neutrinos by the ordinary DW-mechanism~\cite{Dodelson:1993je}, in addition to the production by the mechanism proposed here. This contribution depends on the active-sterile mixing angle $\theta_1$ of the keV sterile neutrino $N_1$, and it can be estimated by the approximate formula~\cite{Kusenko:2009up}:
\begin{equation}
 \Omega_{N_1,\rm DW} h^2 \approx 0.2 \cdot \frac{\sin^2 \theta_1}{3\cdot 10^{-9}} \left( \frac{M_1}{3~{\rm keV}} \right)^{1.8}.
 \label{eq:DW-contr}
\end{equation}
Note that, if the keV sterile neutrino makes up all the DM in the Universe and if it is unstable under $N_1 \to \nu \gamma$, then there is a strong bound from the non-observation of the corresponding X-ray line (see Refs.~\cite{Canetti:2012vf,Canetti:2012kh,Merle:2013ibc} for recent collections of bounds). In the plots, we have represented the corresponding \emph{maximal} (i.e., for the largest allowed value of $\sin^2 \theta_1$) addition of particle production due to the DW mechanism by the purple bands. As can be seen from the plots, this would shift the allowed regions (i.e., the regions where the \emph{total} abundance of keV neutrinos, as produced by both mechanisms together, is within the $3\sigma$ regions of Planck) towards slightly larger values of $m_\sigma$. For very low $M_1$, there is a considerable DW-production resulting from the comparatively weak X-ray bound in this mass region. In this region, nearly all the DM can be produced by the DW-mechanism, which for these masses completely dominates the production by frozen-in scalars, if the maximally possible value is taken for the active-sterile mixing. However, from studies of the Lyman--$\alpha$ forest, the corresponding lower bound on the keV sterile neutrino mass, when the DW-mechanism is at work, is between 8~and 10~keV~\cite{Boyarsky:2008xj} (note that this accidentally coincides with the light red HDM region in our plots). Thus, this region of the parameter space is excluded. On the other hand, depending on the exact value of the active-sterile mixing, the combined abundance of keV sterile neutrinos produced by both mechanisms could also lay in between the orange and purple bands, that we indicate in the plots. We want to stress that the orange bands correspond to production by scalar FIMPs \emph{only}, i.e., this is the region of correct abundance for a vanishing active-sterile mixing, $\theta_1 \equiv 0$. While this may not be desirable from a phenomenological point of view (e.g.\ for a possible detection of the X-ray line~\cite{Abazajian:2006jc,Loewenstein:2009cm,Boyarsky:2010ci} or for a potential detection of modifications of neutrino-less double beta decay~\cite{Merle:2013ibc,Bezrukov:2005mx}), vanishing active-sterile mixing may be very natural in certain settings~\cite{Allison:2012qn,Sierra:2008wj,Gelmini:2009xd,Ma:2012if}. In such frameworks it would be impossible to produce keV sterile neutrinos via the DW and/or SF mechanisms, but our mechanism (as well as the version in which the scalar freezes out) could be easily used as an alternative.

Finally, we have also indicated possible mass limits arising from the X-ray bound. For example, if the bound on the active-sterile mixing is taken to be $\sin^2 (2 \theta_1) < 10^{-13}$, this excludes keV sterile neutrino masses $M_1$ above 64.5~keV, while the values below are consistent with the X-ray bound (but not necessarily with the HDM bound). If active-sterile mixing is not present, then there is no fixed upper bound on $M_1$, and alternative scenarios with stable, e.g., MeV or GeV sterile neutrino DM could be found, too.

The general message of our plots is that there is considerable room for keV sterile neutrinos to be produced by scalar FIMPs and to be compatible with all bounds. Accordingly, if in a certain setting the Higgs portal coupling of a singlet scalar is bound to be very small, it could still be used to produce sterile neutrino DM.

\section{\label{sec:conc}Conclusions}

In this paper, we have presented a new and successful mechanism for the production of keV sterile neutrino DM. The mechanism is based on the so-called freeze-in of a scalar particle $\sigma$, which has too feeble interactions with SM particles (and hence with the primordial thermal plasma) to ever enter thermal equilibrium, but whose mixing with the SM-like Higgs boson is nevertheless large enough to gradually produce a significant abundance of $\sigma$'s in the early Universe. If these $\sigma$'s are unstable, they can decay efficiently into pairs of keV sterile neutrinos $N_1$, thereby generating the required DM abundance. We have numerically solved the corresponding system of coupled Boltzmann equations and we have also presented a discussion of all potentially relevant bounds.

Depending on the exact value of the first generation active-sterile mixing angle, the DM abundance generated by the mechanism proposed here must be corrected by a contribution from the (generic) Dodelson-Widrow mechanism. We have estimated the maximal effect of this additional amount of keV neutrinos, which alters the successful regions in the parameter space, without however spoiling the proposed production mechanism. On the other hand, it is worth to note that our mechanism does not at all rely on active-sterile mixing, and it could very well live even with a vanishing active-sterile mixing angle. This point could be particularly interesting for models which avoid the strong X-ray observational bound on the active-sterile mixing, by stabilising the keV neutrinos and at the same time forbidding any mixture of active and sterile states.

While similar mechanisms had been proposed previously for early and late \emph{freeze-out} (and subsequent decay) of the scalar, our proposal opens up a new window in a region of the parameter space where freeze-out is not at all possible. This is particularly interesting for models which predict a very small Higgs portal coupling between the singlet scalar field $\sigma$ and the SM-like Higgs. Apart from being applicable to many settings where a suitable scalar is available, the main advantage of our production mechanism is that it happens at relatively early times, thereby causing the DM particles to become non-relativistic already at high temperatures (which they do not feel due to their feeble interactions). Hence, depending on the exact values of the parameters, the keV neutrinos can be cold DM in a significant fraction of the parameter space. This is particularly interesting in case the X-ray bound can be circumvented in a concrete model, in which case a significant region of the $M_1$--$m_\sigma$ parameter plane can lead to the correct relic abundance of DM and be consistent with all bounds.

\section*{Acknowledgements}
We would like to thank T.~Hambye, A.~Pukhov, O.~Ruchayskiy, and A.~Schneider for useful discussions. VN acknowledges hospitality at MPIK Heidelberg, where part of this work was completed. AM acknowledges support by a Marie Curie Intra-European Fellowship within the 7th European Community Framework Programme FP7-PEOPLE-2011-IEF, contract PIEF-GA-2011-297557. VN acknowledges support by the Spanish research Grant FPA2010-20807, by the consolider-ingenio 2010 program grants CUP (CSD-2008-00037) and by the CPAN. DS acknowledges support by the International Max Planck Research School for Precision Tests of Fundamental Symmetries. Finally, AM and VN both acknowledge partial support from the European Union FP7 ITN-INVISIBLES (Marie Curie Actions, PITN-GA-2011-289442).

\appendix

\renewcommand{\theequation}{A-\arabic{equation}}
\setcounter{equation}{0}  
 
\section{\label{sec:A_A}Appendix: Effective degrees of freedom}

We followed the notation of Ref.~\cite{Gondolo:1990dk}, where the energy density $\rho_i$ and the entropy density $s_i$ for a particle species $i$ are defined as: 
\begin{eqnarray}
\rho_i\left(T_i\right)&=&g_{\rm eff}^i\left(T_i\right)\frac{\pi^2}{30}T_i^4\,,\\
s_i\left(T_i\right)&=&h_{\rm eff}^i\left(T_i\right)\frac{2\pi^2}{45}T_i^3\,,
\end{eqnarray}
with the temperature $T_i$ of the particle species $i$. The effective degrees of freedom $g_{\rm eff}^i$ and $h_{\rm eff}^i$ for energy and entropy density, respectively, are defined as
\begin{eqnarray}
g_{\rm eff}^i\left(T_i\right) &=& \frac{15g_i}{\pi^4}x_i^4\int\limits_1^\infty dy\,y^2\sqrt{y^2-1}\frac{1}{e^{yx_i}+\eta_i}\,,\\
h_{\rm eff}^i\left(T_i\right) &=& \frac{45g_i}{12\pi^4}x_i^4\int\limits_1^\infty dy\,\left(4y^2-1\right)\sqrt{y^2-1}\frac{1}{e^{yx_i}+\eta_i}\,,
\end{eqnarray}
with $x_i \equiv m_i/T_i$, $\eta_i=1$ for Fermi-Dirac, $\eta_i=-1$ for Bose-Einstein and $\eta_i=0$ for Maxwell-Boltzmann. The number of internal degrees of freedom is denoted by $g_i$.

In our numerics, we accounted for the contribution of the real scalar singlet $\sigma$ and the sterile neutrino $N_1$ to the total energy and entropy effective degrees of freedom given as
\begin{eqnarray}
g_{\rm{ eff}}(T)=\sum_i g_{\rm eff}^i (T_i)\frac{T^4_i}{T^4}\,,\\
h_{\rm{ eff}}(T)=\sum_i h_{\rm eff}^i (T_i)\frac{T^3_i}{T^3}\,.
\end{eqnarray}

\renewcommand{\theequation}{B-\arabic{equation}}
\setcounter{equation}{0}  
  
\section{\label{sec:A_B}Appendix: Modified Bessel functions}

The modified Bessel functions $K_n(x)$ of the second kind obey the identity
\begin{eqnarray}
K_n\left(x\right)=\frac{\sqrt{\pi}}{\left(n-\frac{1}{2}\right)!}\left(\frac{1}{2}x\right)^n \int\limits_1^\infty dy\,\frac{(y^2-1)^{n-\frac{1}{2}}}{e^{xy}}\,.
\end{eqnarray} 
For a Maxwell Boltzmann distribution with zero chemical potential, the equilibrium number density $n_{\rm eq}$ of a particle with $g$ internal degrees of freedom  and mass $m$ is
\begin{eqnarray}
n_{\rm eq}&=&\frac{g}{(2\pi)^3}\int\limits_0^\infty d^3p\ e^{-E/T}=\frac{g}{2\pi^2}\int\limits_{m}^\infty dE\ E\sqrt{E^2-m^2}\ e^{-E/T}=
\nonumber\\
&=&m^3\frac{g}{2\pi^2}\frac{1}{x}K_2(x)\,,
\end{eqnarray}
where $x=m/T$. For the yield $Y=\frac{n}{s}$ with entropy density $s=\frac{2\pi^2}{45}h_{\rm{ eff}}T^3$ follows:
\begin{eqnarray}
Y_{\rm{eq}}=\frac{45g}{4\pi^4}\frac{x^2}{h_{\rm{eff}}}K_2(x)\,.
\end{eqnarray}

\renewcommand{\theequation}{C-\arabic{equation}}
\setcounter{equation}{0}  

\section{\label{sec:A_C}Appendix: Annihilation and decay reactions}

In the usual Friedman-Robertson-Walker metric, the Boltzmann equation for the number density $n$ of a particle species can be written as
\begin{eqnarray}
\frac{d}{dt}n+3Hn=C[n]\,,
\end{eqnarray}
where $C$ is the collision operator expressing the number of particles per phase space volume that are lost or gained per unit time due to interactions with other particles. 

For the standard annihilaton process $\sigma\,\sigma\rightleftharpoons {\rm SM}\,\ {\rm SM}$ of a real scalar singlets $\sigma$ into Standard Model particles, the Boltzmann equation for the number density $n_\sigma$ reads
\begin{eqnarray}\label{Boltzmann:n_A}
\frac{d}{dt}n_\sigma+3Hn_\sigma=-\langle\sigma_{\rm ann} v\rangle (n_\sigma^2-n^2_{\sigma,\rm{eq}}) 
\simeq \langle\sigma_{\rm ann} v\rangle n^2_{\sigma,\rm{eq}}\,,
\end{eqnarray}
where the latter approximation is valid for the freeze-in case, for which the initial number density and thus the initial abundance can be neglected~\cite{Hall:2009bx}. Furthermore, $\langle\sigma_{\rm ann} v\rangle$ is the relativistic thermally averaged annihilation cross section and $v$ is the M{\o}ller velocity. Following the discussion of Ref.~\cite{Gondolo:1990dk}, it is possible to write
\begin{eqnarray}\label{thermAverFin}
\langle\sigma_{\rm ann} v\rangle=\frac{1}{8 m_\sigma^4 T K_2^2 (m_\sigma/T)}\int\limits_{4 m_\sigma^2}^\infty ds\ \sigma_{\rm ann} (s-4m_\sigma^2)\sqrt{s}\ K_1 \left(\frac{\sqrt{s}}{T}\right)\,.
\end{eqnarray}
We have generated the correct Feynman rules using LanHEP~\cite{Semenov:1996es} and we have used micrOMEGAs~\cite{Belanger:2010gh} for the calculation of Eq.~\eqref{thermAverFin}.

In the radiation dominated era, the Hubble expansion rate can be expressed as
\begin{equation}
H=\sqrt{\frac{4\pi^3G_Ng_{\rm{ eff}}}{45}}T^2\,,
\end{equation}
where $G_N$ is Newton's gravitational constant. Furthermore, in the radiation dominated era, the expansion age $t$ of the Universe with $\Omega_{\rm{tot}}=1$ equals:
\begin{equation}
t=\frac{1}{2 H}\,.
\end{equation}
In terms of the abundance $Y=\frac{n}{s}$ with the entropy density $s=\frac{2\pi^2}{45}h_{\rm{ eff}}T^3$, it follows:
\begin{equation}
\frac{d}{dT}Y_\sigma^\mathcal{A} = -\sqrt{\frac{\pi}{45G_N}}\sqrt{g_\ast}\langle\sigma_{\rm ann} v\rangle 
Y^2_{\sigma,\rm{eq}}\label{yield:A}\,,
\end{equation}
with the definition
\begin{equation}
\sqrt{g_\ast}\equiv\frac{h_{\rm{ eff}}}{\sqrt{g_{\rm{ eff}}}}\left(1 + \frac{1}{3}\frac{T}{h_{\rm{ eff}}}\frac{dh_{\rm{ eff}}}{T} \right)\,.
\end{equation}
The superscript $\mathcal{A}$ serves as indication of the annihilation process.

The decay processes $\sigma\rightarrow N_1 N_1$ of a real scalar singlet $\sigma$ into two sterile neutrinos $N_1$ is described by the following phase space integration:
\begin{equation}
\int\frac{d^3p_\sigma}{(2\pi)^3 2E_\sigma}\frac{d^3 p_{N_1}}{(2\pi)^3 2 E_{N_1}}\frac{d^3 p'_{N_1}}{(2\pi)^3 2 E'_{N_1}}(2\pi)^4\delta^{(4)}(p_{N_1}+p'_{N_1}-p_\sigma)|\mathcal{M}|^2_{\sigma\rightarrow N_1N_1}f_\sigma(1-f_{N_1})(1-f'_{N_1}).
\end{equation}
Neglecting, the Pauli blocking and enhancing factors, we can define
\begin{equation}\label{eq:decay1}
\int\frac{d^3 p_{N_1}}{(2\pi)^3 2E_{N_1}}\frac{d^3 p'_{N_1}}{(2\pi)^3 2E'_{N_1}}(2\pi)^4\delta^{(4)}(p_{N_1}+p'_{N_1}-p_\sigma)|\mathcal{M}|^2_{\sigma\rightarrow N_1N_1}
\equiv 2E_\sigma \Gamma^\ast(\sigma \rightarrow N N)\,,
\end{equation}
with $\Gamma^\ast(\sigma \rightarrow N_1 N_1)$ the decay width for the particle at energy $E_\sigma$. The above phase space integration yields:
\begin{equation}
\int dn_\sigma\,\Gamma^\ast(\sigma \rightarrow N_1 N_1)= n_\sigma\,\langle \Gamma(\sigma \rightarrow N_1 N_1)\rangle\,, \ \ \ {\rm where}
\end{equation}
\begin{equation}\label{eq:decay2}
\langle\Gamma(\sigma \rightarrow N_1 N_1)\rangle=\frac{\int d^3p_\sigma\Gamma^\ast(\sigma \rightarrow N_1 N_1)e^{- E_\sigma/T}}{\int d^3p_\sigma e^{-E_\sigma/T}}= \frac{K_1(x)}{K_2(x)}\Gamma(\sigma \rightarrow N_1 N_1)\,,
\end{equation}
with $\Gamma(\sigma \rightarrow N_1 N_1)$ the decay width in the rest frame of the decaying particle $\sigma$, i.e., $\Gamma(\sigma \rightarrow N_1 N_1)=\frac{E_\sigma}{m_\sigma}\Gamma^\ast(\sigma \rightarrow N_1 N_1)$. Thus, for the decay process $\sigma\rightarrow N_1N_1$ of a real scalar singlet $\sigma$ into two sterile neutrinos $N_1$, the Boltzmann equation for the number density $n_{N_1}$ reads as (again we define $x = m_\sigma / T$)
\begin{equation}\label{Boltzmann:n_D}
\frac{d}{dt}n_{N_1}+3Hn_{N_1}=2\frac{K_1(x)}{K_2(x)}\Gamma(\sigma \rightarrow N_1 N_1)\,n_\sigma\,.
\end{equation}
The factor $2$ accounts for the fact that two sterile neutrinos $N_1$ are produced per decay. In terms of the abundance $Y=\frac{n}{s}$ with the entropy density $s=\frac{2\pi^2}{45}h_{\rm{ eff}}T^3$, it follows:
\begin{equation}\label{yield:D}
\frac{d}{dT}Y_{N_1}^{\mathcal{D}}=-\sqrt{\frac{45}{\pi^3G_N}}\frac{1}{T^3}\frac{1}{\sqrt{g_{\rm eff}}}\frac{K_1(x)}{K_2(x)}\Gamma(\sigma \rightarrow N_1 N_1)\,Y_\sigma\,,
\end{equation}
where the superscript $\mathcal{D}$ serves as indication of the decay process.

\renewcommand{\theequation}{D-\arabic{equation}}
\setcounter{equation}{0}  

\section{\label{sec:A_D}Appendix: Free-streaming horizon}

With the distribution function $f(p,t) $ of the DM particle, given as
\begin{equation}
 f(p,t) = \frac{\beta}{p/T_{\rm WDM}} \exp \left(-\frac{p^2}{T_{\rm WDM}^2} \right),
\end{equation}
cf.\ Eq.~\eqref{eq:FSdist_1}, its average momentum $ \langle p(t) \rangle $ equals
\begin{equation}
 \langle p(t) \rangle = \frac{\int d^3 p\ p\ f(p,t)}{\int d^3 p\ f(p,t)} = \frac{\int_{p=0}^{\infty} d p\ p^2\ e^{-p^2/T_{\rm WDM}^2}}{\int_{p=0}^{\infty} d p\ p\  e^{-p^2/T_{\rm WDM}^2}}\,,
 \label{eq:FSdist_6}
\end{equation}
which determines the non-relativistic average velocity $\langle v(t) \rangle = \langle p(t) \rangle/M_1$ of the DM particle with mass $M_1$.

The free streaming horizon $r_{\rm FS}$ can then be calculated as, cf.\ Eq.~\eqref{eq:FS-horizon},
\begin{equation}
 r_{\rm FS} = \int\limits_{t_{\rm in}}^{t_0} \frac{\langle v(t) \rangle}{a(t)} dt,
 \label{eq:FS-horizon_app}
\end{equation}
where $t_{\rm in}$ is the initial time at which the integration starts, $t_0$ is the current time, $\langle v(t) \rangle$ is the mean velocity of the DM particles, and $a(t)$ is the scale factor.

The non-relativistic transition time $t_{\rm nr}$ is defined by $\langle p(t_{\rm nr}) \rangle = M_1$. For an early non-relativistic transition, i.e., the DM particle becomes non-relativistic at $t_{\rm nr} < t_{\rm eq}$, where $t_{\rm eq}$ is the time of matter-radiation equality, the integral in Eq.~\eqref{eq:FS-horizon_app} can be split into three:
\begin{eqnarray}
 r_{\rm FS}&=& \int\limits_{t_{\rm in}}^{t_0} \frac{\langle v(t) \rangle}{a(t)} dt = \int\limits_{t_{\rm in}}^{t_{\rm nr}}\frac{dt}{a(t)}+\int\limits_{t_{\rm nr}}^{t_{\rm eq}}\frac{\langle v(t) \rangle}{a(t)} dt+\int\limits_{t_{\rm eq}}^{t_0}\frac{\langle v(t) \rangle}{a(t)} dt\nonumber\\
 &\simeq&\frac{2 \sqrt{t_{\rm eq} t_{\rm nr}}}{a_{\rm eq}}+\frac{\sqrt{t_{\rm eq} t_{\rm nr}}}{a_{\rm eq}} \ln \left( \frac{t_{\rm eq}}{t_{\rm nr}} \right)+\frac{3 \sqrt{t_{\rm eq} t_{\rm nr}}}{a_{\rm eq}} = \frac{\sqrt{t_{\rm eq} t_{\rm nr}}}{a_{\rm eq}} \left[ 5 + \ln \left( \frac{t_{\rm eq}}{t_{\rm nr}} \right) \right]\,. \label{eq:FS-horizon_early}
\end{eqnarray}
In the case of a late transition, i.e., $t_{\rm nr}>t_{\rm eq}$, it follows instead:
\begin{eqnarray}
 r_{\rm FS}&=& \int\limits_{t_{\rm in}}^{t_0} \frac{\langle v(t) \rangle}{a(t)} dt  = \int\limits_{t_{\rm in}}^{t_{\rm eq}}\frac{dt}{a(t)}+\int\limits_{t_{\rm eq}}^{t_{\rm nr}}\frac{\langle v(t) \rangle}{a(t)} dt+\int\limits_{t_{\rm nr}}^{t_0}\frac{\langle v(t) \rangle}{a(t)} dt\nonumber\\
 &\simeq&\frac{2 t_{\rm eq}}{a_{\rm eq}}+\left( \frac{3 t_{\rm eq}^{2/3} t_{\rm nr}^{1/3}}{a_{\rm eq}} - \frac{3 t_{\rm eq}}{a_{\rm eq}}\right)+\frac{\sqrt{\pi}}{2} \frac{m_\sigma/2}{M_1} \sqrt{\frac{t_{\rm in}}{t_{\rm eq}}} \frac{3\ t_{\rm eq}^{4/3} }{a_{\rm eq} t_{\rm nr}^{1/3}}\nonumber\\
 &=& \frac{3 t_{\rm eq}^{2/3} t_{\rm nr}^{1/3}}{a_{\rm eq}} - \frac{t_{\rm eq}}{a_{\rm eq}} + \frac{\sqrt{\pi}}{2} \frac{m_\sigma/2}{M_1} \sqrt{\frac{t_{\rm in}}{t_{\rm eq}}} \frac{3\ t_{\rm eq}^{4/3} }{a_{\rm eq} t_{\rm nr}^{1/3}}\,. 
 \label{eq:FS-horizon_late}
\end{eqnarray}
Note that both expressions, Eqs.~\eqref{eq:FS-horizon_early} and~\eqref{eq:FS-horizon_late}, exactly coincide in the limit $t_{\rm nr} \to t_{\rm eq}$.


 \bibliographystyle{apsrev}
\bibliography{fimp}

\end{document}